\documentclass{ieeeaccess}
\usepackage{cite}
\usepackage{amsmath,amssymb,amsfonts}
\usepackage{graphicx}
\usepackage{textcomp}
\usepackage{empheq}
\usepackage{algorithmicx}
\usepackage{algorithm,algpseudocode}
\usepackage{setspace}
\usepackage{subcaption}
\usepackage{pifont}
\usepackage[table]{xcolor}

\newcommand{\cmark}{\ding{51}}%
\newcommand{\xmark}{\ding{55}}%

\usepackage{float}
\usepackage{url}
\usepackage{dblfloatfix} 
\usepackage{mathtools}
\usepackage{multirow}

\usepackage{bm}
\makeatletter
\AtBeginDocument{\DeclareMathVersion{bold}
\SetSymbolFont{operators}{bold}{T1}{times}{b}{n}
\SetSymbolFont{NewLetters}{bold}{T1}{times}{b}{it}
\SetMathAlphabet{\mathrm}{bold}{T1}{times}{b}{n}
\SetMathAlphabet{\mathit}{bold}{T1}{times}{b}{it}
\SetMathAlphabet{\mathbf}{bold}{T1}{times}{b}{n}
\SetMathAlphabet{\mathtt}{bold}{OT1}{pcr}{b}{n}
\SetSymbolFont{symbols}{bold}{OMS}{cmsy}{b}{n}
\renewcommand\boldmath{\@nomath\boldmath\mathversion{bold}}}
\makeatother

\def\BibTeX{{\rm B\kern-.05em{\sc i\kern-.025em b}\kern-.08em
    T\kern-.1667em\lower.7ex\hbox{E}\kern-.125emX}}

\begin{document}
\doi{10.1109/ACCESS.2024.3476410}

\title{Optimizing Energy-Harvesting Hybrid VLC/RF Networks with Random Receiver Orientation}
\author{\uppercase{Amir~Hossein~Fahim~Raouf}\authorrefmark{1}, \IEEEmembership{Graduate Student Member,~IEEE},
\uppercase{Chethan~Kumar~Anjinappa}\authorrefmark{2}, and Ismail~Guvenc\authorrefmark{1},
\IEEEmembership{Fellow,~IEEE}}

\address[1]{A. H. F. Raouf and I. Guvenc are with the Department of Electrical and Computer Engineering,
North Carolina State University, Raleigh, NC (e-mail: amirh.fraouf@ieee.org and iguvenc@ncsu.edu)}
\address[2]{C. K. Anjinappa is with the Ericsson Research, Santa Clara, CA 95054, USA (email: chethan.anjinappa@ericsson.com)}
\tfootnote{This paper is presented in part at IEEE Globecom Workshops (GC Wkshps), December 2022~\cite{raouf2022optimal}. This work is supported in part by NSF under the grant CNS-1910153.}

\markboth
{Author \headeretal: }
{Author \headeretal: }

\corresp{Corresponding author: A. H. F. Raouf (e-mail: amirh.fraouf@ieee.org).}

\begin{abstract}
This paper investigates an indoor hybrid visible light communication (VLC) and radio frequency (RF) scenario with two-hop downlink transmission. A light emitting diode (LED) transmits both data and energy via VLC to an energy-harvesting relay node, which then uses the harvested energy to retransmit the decoded information to an RF user in the second phase. The design parameters include the direct current (DC) bias and the time allocation for VLC transmission. We formulate an optimization problem to maximize the data rate under decode-and-forward relaying with fixed receiver orientation. The non-convex problem is decomposed into two sub-problems, solved iteratively by fixing one parameter while optimizing the other. Additionally, we analyze the impact of random receiver orientation on the data rate, deriving closed-form expressions for both VLC and RF rates. An exhaustive search approach is employed to solve the optimization, demonstrating that joint optimization of DC bias and time allocation significantly enhances the data rate compared to optimizing DC bias alone.
\end{abstract}


\titlepgskip=-21pt

\maketitle

\section{Introduction}\label{sec:intro}
The burgeoning demand for wireless communication services and emerging technologies has significantly strained the radio frequency (RF) spectra~\cite{matheus2019visible}. This strain has led to substantial challenges in spectrum management, particularly in dense environments such as conference halls, stadiums, shopping centers, and airports, where RF resources are increasingly scarce. Visible light communication (VLC) has emerged as a promising complementary technology to RF-based wireless systems, offering the potential to offload users from congested RF bands while simultaneously providing illumination~\cite{burchardt2014vlc}. However, despite its promise, VLC systems face several limitations, including coverage constraints due to line-of-sight (LoS) requirements and susceptibility to environmental interference, which limit their standalone effectiveness in real-world scenarios.

\begin{figure*}[t]
\centering
\includegraphics[trim=0.1cm 0.1cm 0.1cm 0.3cm, clip,width=0.7\linewidth]{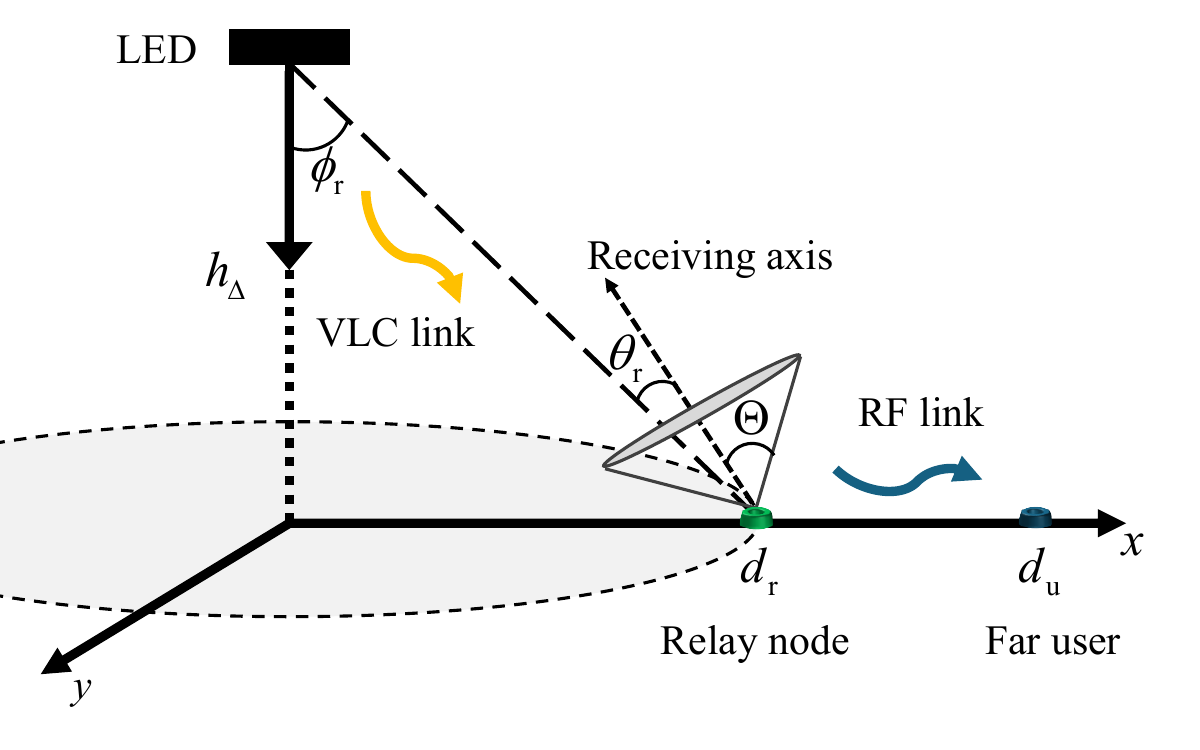}
\caption{The system model for the considered VLC-RF transmission scenario. The VLC link carries both data and energy to the relay node. The harvested energy is then used at the relay node to forward the data to the far RF user.}
\label{fig:system_model}
\end{figure*}

To address these challenges, recent research has focused on hybrid VLC-RF systems designed to leverage the strengths of both technologies. These systems can achieve high-speed data transmission through VLC links, whereas RF links provide seamless coverage and overcome VLC's LoS and mobility constraints~\cite{basnayaka2015hybrid}. Despite these advantages, hybrid VLC-RF systems present new challenges, particularly for indoor applications such as the Internet of Things (IoT) and wireless sensor networks~\cite{delgado2020hybrid, pan2019simultaneous}. A critical bottleneck in these networks is the power constraint, as devices often operate with limited energy resources.
One promising solution to this challenge is the incorporation of energy harvesting (EH) techniques, which allow devices to scavenge energy from the surrounding environment, reducing reliance on battery power and improving network sustainability, as illustrated in Fig.~\ref{fig:system_model}.

\begin{figure*}[!t]
\centering
\includegraphics[width=0.7\linewidth]{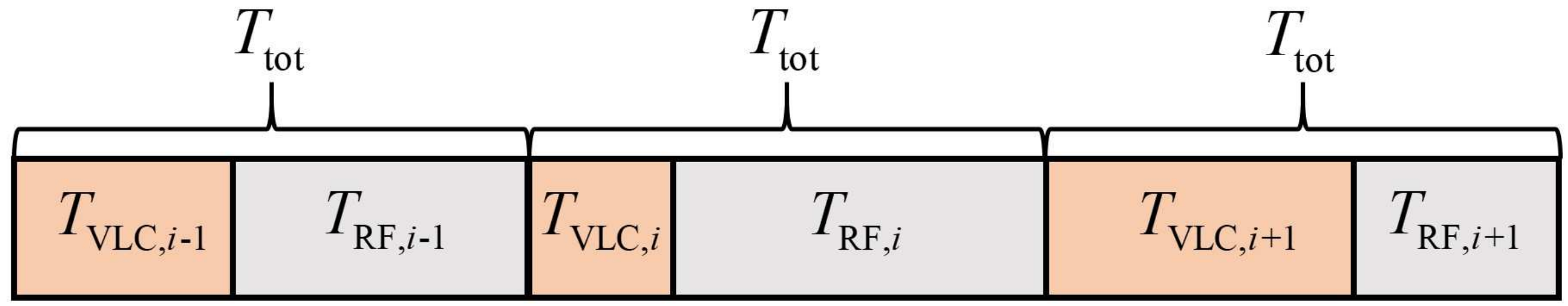}
\caption{The transmission block under consideration with consecutive time periods dedicated for VLC (relay) and RF (access) links. The VLC link is used both as a backhaul to relay the data and for energy harvesting.}
\label{fig:transmission_block}
\end{figure*}

Existing literature on energy-harvesting hybrid VLC-RF systems predominantly focuses on optimizing the direct current (DC) bias to either maximize the data rate or minimize the outage probability~\cite{rakia2016optimal,yapici2020energy,pan2017secure,zhang2018cooperative,pan20193,peng2020performance,peng2021end,zhang2021slipt,zargari2021resource,xiao2021cooperative,rallis2023rsma,tran2019ultra,guo2021achievable, ghosh2022performance}. To the best of our knowledge, the optimization of VLC and RF resources for hybrid RF-VLC links for a multi-hop scenario, as shown in Fig.~\ref{fig:system_model}, remains
unexplored. In this paper, we investigate the performance of EH for an indoor hybrid VLC-RF scenario. In particular, we allocate a portion of each transmission block to VLC and the rest to RF transmission in an adaptive manner. The light emitting diode (LED) transmits both data and energy to a relay node with energy harvesting capability in the first phase as illustrated in Fig.~\ref{fig:transmission_block} (i.e., VLC transmission). During the second phase (RF communication), the relay transmits the decoded information to the distant RF user using the harvested energy. Also, during this phase, the~LED continues to transmit power (no information) to the relay node, aiming to harvest energy that can be utilized by the~RF relay in the next transmission block. The key contributions of this work are summarized as follows.

\begin{table*}[t!]
\caption{Summary of the most relevant works and their contributions.}
\label{table:Lit_summ}
\begin{center}
\scalebox{0.71}{
\begin{tabular}{|c|l|l|l|c|c|l|l|c|}
\hline
\cellcolor{gray!25} \textbf{Ref.}                             & \cellcolor{gray!25}{\begin{tabular}[c]{@{}l@{}}\textbf{Wireless}\\ \textbf{Technology}\end{tabular}} & \cellcolor{gray!25}\textbf{Scheme} & \cellcolor{gray!25}{\begin{tabular}[c]{@{}l@{}}\textbf{Relay}\\ \textbf{Power}\end{tabular}} & \cellcolor{gray!25}{\begin{tabular}[c]{@{}c@{}}\textbf{EH during}\\ \textbf{VLC Transmission}\end{tabular}} & \cellcolor{gray!25}{\begin{tabular}[c]{@{}c@{}}\textbf{EH during}\\ \textbf{RF Transmission}\end{tabular}} & \cellcolor{gray!25}\textbf{Performance Metric}                                                     & \cellcolor{gray!25}\textbf{Optimization Parameters}                                                     &\cellcolor{gray!25} {\begin{tabular}[c]{@{}c@{}}\textbf{RO}\\ \end{tabular}} \\ \hline
\cite{rakia2016optimal}                          & Hybrid VLC-RF                                                          & Dual-hop        & Harvested energy                                               & \cellcolor{green!25}\cmark                                                                   & \cellcolor{green!25}\cmark                                                                 & Data rate                                                                       & DC bias                                                                              & \cellcolor{red!25}\xmark                                                                 \\ \hline
\cite{yapici2020energy}                          & Hybrid VLC-RF                                                          & Dual-hop        & Harvested energy                                               & \cellcolor{green!25}\cmark                                                                  & \cellcolor{red!25}\xmark                                                                 & \begin{tabular}[c]{@{}l@{}}Energy efficiency\\ Spectral efficiency\end{tabular} & No parameter optimization                                                                                    & \cellcolor{red!25}\xmark                                                                \\ \hline
\cite{pan2017secure}                             & Hybrid VLC-RF                                                          & Dual-hop & Harvested energy                                               & \cellcolor{green!25}\cmark                                                                  & \cellcolor{red!25}\xmark                                                                 & Secrecy outage probability                                                      & No parameter optimization                                                                                    & \cellcolor{red!25}\xmark                                                                \\ \hline
\cite{zhang2018cooperative}                      & Hybrid VLC-RF                                                          & Cooperative     & External                                                       & \cellcolor{red!25}\xmark                                                                  & \cellcolor{red!25}\xmark                                                                 & \begin{tabular}[c]{@{}l@{}}Outage probability\\ Symbol error rate\end{tabular}  & No parameter optimization                                                                                    & \cellcolor{red!25}\xmark                                                                \\ \hline
\cite{pan20193}                                  & Hybrid VLC-RF                                                          & Dual-hop & Harvested energy                                               & \cellcolor{green!25}\cmark                                                                  & \cellcolor{red!25}\xmark                                                                 & Outage probability                                                              & No parameter optimization                                                                                    & \cellcolor{red!25}\xmark                                                                \\ \hline
\cite{peng2020performance}                       & Hybrid VLC-RF                                                          & Dual-hop        & Harvested energy                                               & \cellcolor{green!25}\cmark                                                                  & \cellcolor{red!25}\xmark                                                                 & Outage probability                                                              & No parameter optimization                                                                                    & \cellcolor{red!25}\xmark                                                                \\ \hline
\cite{peng2021end}                               & Hybrid VLC-RF                                                          & Dual-hop        & Harvested energy                                               & \cellcolor{green!25}\cmark                                                                  & \cellcolor{red!25}\xmark                                                                 & Outage probability                                                              & \begin{tabular}[c]{@{}l@{}}DC bias\\ Peak amplitude\end{tabular}                     & \cellcolor{red!25}\xmark                                                                \\ \hline
\cite{zhang2021slipt}                            & Hybrid VLC-RF                                                          & Dual-hop        & Harvested energy                                               & \cellcolor{green!25}\cmark                                                                  & \cellcolor{red!25}\xmark                                                                 & Outage probability                                                              & No parameter optimization                                                                                    & \cellcolor{red!25}\xmark                                                                \\ \hline
\cite{zargari2021resource}                       & Hybrid VLC-RF                                                          & Dual-hop        & Harvested energy                                               & \cellcolor{green!25}\cmark                                                                  & \cellcolor{red!25}\xmark                                                                 & Data rate                                                                       & \begin{tabular}[c]{@{}l@{}}LED transmit power\\ UL/DL transmission time\end{tabular} & \cellcolor{red!25}\xmark                                                                \\ \hline
\cite{xiao2021cooperative} & Hybrid VLC-RF                                                          & Dual-hop        & Harvested energy                                               & \cellcolor{green!25}\cmark                                                                  & \cellcolor{red!25}\xmark                                                                 & Outage probability                                                                       & DC bias                                                                                    & \cellcolor{red!25}\xmark                                                                \\ \hline
\cite{rallis2023rsma}      & Hybrid VLC-RF                                                          & Cooperative     & External                                                       & \cellcolor{red!25}\xmark                                                                  & \cellcolor{red!25}\xmark                                                                 & Data rate                                                                       & \begin{tabular}[c]{@{}l@{}}Time duration\\ Power allocation\end{tabular}             & \cellcolor{red!25}\xmark                                                                \\ \hline
\cite{tran2019ultra}                             & Hybrid VLC-RF                                                          & Collaborative   & N/A                                                              & \cellcolor{green!25}\cmark                                                                  & \cellcolor{green!25}\cmark                                                                 & SNR                                                                             & \begin{tabular}[c]{@{}l@{}}DC bias\\ RF beamformers\end{tabular}                     & \cellcolor{red!25}\xmark                                                                \\ \hline
\cite{guo2021achievable}   & Hybrid VLC-RF                                                          & Cooperative     & External                                                       & \cellcolor{green!25}\cmark                                                                  & \cellcolor{red!25}\xmark                                                                 & Data rate                                                                       & \begin{tabular}[c]{@{}l@{}}Access mode selection\\ DC bias\\ Power allocation\end{tabular} & \cellcolor{red!25}\xmark                                                                \\ \hline
\cite{erouglu2018impact}                         & VLC only                                                                    & Direct link               & N/A                                                             & \cellcolor{red!25}\xmark                                                                  & \cellcolor{red!25}\xmark                                                                 & Bit error rate                                                                  & No parameter optimization                                                                                    & \cellcolor{green!25}\cmark                                                                \\ \hline
\cite{fu2021realistic}                           & VLC only                                                                   & Direct link               & N/A                                                             & \cellcolor{red!25}\xmark                                                                  & \cellcolor{red!25}\xmark                                                                 & Outage probability                                                              & No parameter optimization                                                                                    & \cellcolor{green!25}\cmark                                                                \\ \hline
\cite{soltani2019impact}                         & VLC only                                                                   & Direct link               & N/A                                                              & \cellcolor{red!25}\xmark                                                                  & \cellcolor{red!25}\xmark                                                                 & Bit error rate                                                                  & No parameter optimization                                                                                    & \cellcolor{green!25}\cmark                                                                \\ \hline
\cite{soltani2018modeling}                       & VLC only                                                                   & Direct link               & N/A                                                              & \cellcolor{red!25}\xmark                                                                  & \cellcolor{red!25}\xmark                                                                 & SNR                                                                             & No parameter optimization                                                                                    & \cellcolor{green!25}\cmark                                                                \\ \hline
\cite{purwita2018impact}                         & VLC only                                                                   & Direct link               & N/A                                                              & \cellcolor{red!25}\xmark                                                                  & \cellcolor{red!25}\xmark                                                                 & SNR                                                                             & No parameter optimization                                                                                    & \cellcolor{green!25}\cmark                                                                \\ \hline
\begin{tabular}[c]{@{}l@{}} \textbf{This}\\ \textbf{work}\end{tabular}                        & Hybrid VLC-RF                                                          & Dual-hop        & Harvested energy                                               & \cellcolor{green!25}\cmark                                                                  & \cellcolor{green!25}\cmark                                                                 & Data rate                                                                       & \begin{tabular}[c]{@{}l@{}}DC bias\\ Time allocation\end{tabular}                    & \cellcolor{green!25}\cmark                                                                \\ \hline
\end{tabular}
}
\end{center}
\end{table*}

\begin{itemize}

   \item In a related study~\cite{1}, a comparable policy was introduced for a single indoor link that can be based either on~VLC or infrared communications~(IRC) with the aim of maximizing the harvested energy; however, no RF links or relays are considered, nor is the goal to maximize the data rate. In addition, due to existence of a relay in the considered system model, its relative distance to the RF user, and its random orientation, we \textit{dynamically} allocate a portion of each transmission block to VLC and the rest to RF transmission.

    \item For this specific scenario, we formulate an optimization problem for maximizing the data rate at the far user. In particular, different than any existing work in the literature (see e.g.,~\cite{rakia2016optimal}), we incorporate the assigned time duration to VLC link as the design parameters, in addition to the DC bias. We split the joint non-convex optimization problem over these two parameters into two sub-problems and solve them cyclically. First, we fix the assigned time duration for VLC transmission and solve the non-convex problem for DC bias by employing the majorization-minimization~(MM) procedure~\cite{Yusuf_VLC} and~\cite{sun2016majorization}. The second step involves fixing the DC bias obtained from the previous step and solving an optimization problem for the assigned time duration of the VLC~link. 

     \item To the best of our knowledge, this is the first paper that attempts to investigate the effect of random receiver orientation for the relay on the achievable data rate for a hybrid VLC-RF network.
     Unlike the conventional RF wireless networks, the orientation of devices has a significant impact on VLC channel gain, especially for mobile users. Determining the exact information rate is a formidable task and may not offer valuable insights for optimizing the system's information rate. As an alternative approach, we formulate the \textit{average} information rate for the VLC link and the harvested energy based on the orientation distribution. To gain a better understanding of the influence of system and channel parameters, we assume that receiver orientation follows a uniform distribution. From this assumption, we derive a closed-form expression for the lower bound on the average information rate of both VLC and RF. To verify our analysis, we present the results for the VLC and RF information rate using three methods; i.e., exact integral expressions, simulations and the derived closed-form expressions. Based on the obtained closed-form expressions, we find the optimal values of DC bias and time allocation for the system model under consideration. Due to the complexity of the problem, an exhaustive search is conducted to solve it.
\end{itemize}

The remainder of this paper is organized as follows. Section~\ref{sec:lit_rev} presents the literature review. In Section~\ref{sec:sys_model}, we describe our system model. The optimization framework is introduced in Section \ref{sec:opt_fw}, while the optimization problem and our approach to solve it are provided in Section~\ref{sec:sol_app}. In Section \ref{sec:random_or}, we derive the closed-form expressions for both VLC and RF data rate by considering random orientation (RO), numerical results are presented in Section~\ref{sec:num_results}, and finally, the conclusion and future works are suggested in Section~\ref{sec:conclusion}.

\section{Literature Review}\label{sec:lit_rev}
There have been some recent studies on enabling EH for a dual-hop hybrid VLC-RF communication system where the relay can harness energy from a VLC link (first hop), for re-transmitting the data to the end user over the RF link (second hop). For example, Rakia \textit{et al.} in~\cite{rakia2016optimal}
 introduce an optimal design that maximizes the data rate with respect to the DC bias by allocating equal time portions for VLC and RF transmissions.
In another work, Yapici and Guvenc in~\cite{yapici2020energy} investigate the trade-off between energy and spectral efficiency by considering LED power consumption, highlighting the necessity of DC bias optimization.

Using stochastic geometry, in~\cite{pan2017secure} secrecy outage probability and the statistical characteristics of the received signal-to-noise ratio (SNR) are derived in the presence of an eavesdropper for a hybrid VLC-RF system.
Outage probability and symbol error rate are studied in~\cite{zhang2018cooperative} under the assumption that the relay and destination locations are random. They consider both decode-and-forward~(DF) and amplify-and-forward~(AF) schemes and derive the approximated analytical and asymptotic expressions for the outage probability. In~\cite{pan20193}, the outage performance of an IoT hybrid RF-VLC system is investigated where the VLC is considered as the downlink from the LED to the IoT devices, while RF utilizes a non-orthogonal multiple access~(NOMA) scheme for the uplink. Specifically, they report the approximated analytical expressions for the outage probability by utilizing a stochastic geometry approach to model the location and number of terminals in a 3-D room.

Peng \textit{et al.}~\cite{peng2020performance} consider a mobile relay to facilitate communications between the source and destination. They analytically obtain the system's end-to-end outage probability and compare it with simulation results. In a subsequent study, Peng \textit{et al.} in~\cite{peng2021end} extend this work by addressing the minimization of end-to-end outage probability under both average and peak power constraints of the LED source.
Zhang \textit{et al.}~\cite{zhang2021slipt} select the relay from multiple IoT devices randomly distributed within the coverage area of the source. Utilizing channel state information~(CSI), they employ an analytical approach to determine the end-to-end outage probability for two different transmission schemes; without CSI and with statistical CSI.
Zargari \textit{et al.}~\cite{zargari2021resource} investigate the problem of maximizing the sum throughput of multiple users in a hybrid VLC-RF communication system, where users harvest energy during downlink for transmission in uplink.

Peng \textit{et al.}~\cite{xiao2021cooperative} consider a cooperative hybrid VLC-RF relaying network and calculate the outage probability for both VLC and RF users. Furthermore, they derive a sub-optimal DC bias that effectively minimizes the outage probability for the RF user.
Rallis \textit{et al.}~\cite{rallis2023rsma} propose a hybrid VLC-RF network where a VLC access point~(AP) serves two user equipments (UEs), which also function as RF relays to extend network coverage to a third user beyond the VLC cell. Inspired by rate-splitting multiple access, the proposed protocol aims to maximize the weighted minimum achievable rate in the system.
Tran \textit{et al.}~\cite{tran2019ultra} introduce a hybrid VLC-RF ultra-small network where optical transmitters deliver both lightwave information and energy signals, while a multiple-antenna RF AP is employed to transfer wireless power via RF signals.
Guo \textit{et al.}~\cite{guo2021achievable} consider two types of users: information users and EH users. The information users receive data from the LED AP through a time-division multiple access~(TDMA) scheme using either a single-hop VLC-only mode or a relay-assisted dual-hop VLC-RF mode, where the relay has access to an external power source. Utilizing harvesting energy from different energy sources (from visible light and RF signal), Ghosh and Alouini in~\cite{ghosh2022performance} derive the closed-form outage expressions of both two-way licensed user and two-way IoT communications using the DF relaying scheme.

The existing literature on hybrid VLC-RF communication systems primarily assumes that the receiver is fixed and oriented vertically upward, with the effect of random receiver orientation on such systems yet to be extensively reported. Receiver orientation significantly impacts the availability of LoS links in VLC networks. Ero{\u{g}}lu \textit{et al.}~\cite{erouglu2018impact} present the statistical distribution of the VLC channel gain in the presence of random orientation for mobile users. Fu \textit{et al.}~\cite{fu2021realistic} derive the average channel capacity and outage probability based on the statistical characteristics of the channel when VLC receivers have random locations and orientations. Rodoplu \textit{et al.}~\cite{rodoplu2020characterization} study the behavior of human users and LoS availability in an indoor environment. They further derive the outage probability and analyze the effect of random orientation on inter-symbol interference. Utilizing the Laplace distribution, Soltani \textit{et al.}~\cite{soltani2019impact} derive the probability density function of SNR and bit error rate for an indoor scenario. Recent efforts have also been made on experimental measurements to model receiver orientation~\cite{soltani2018modeling, purwita2018impact}. Table~\ref{table:Lit_summ} summarizes recent studies on hybrid VLC-RF communication systems and VLC receiver orientation, comparing them with our current work.

\section{System Model}\label{sec:sys_model}
Fig.~\ref{fig:system_model} illustrates the hybrid VLC-RF system under consideration. We assume a relay equipped with a single photo-detector~(PD), energy-harvesting circuitry, and a transmit antenna for RF communications. The relay is located at vertical and horizontal distances, $h_\Delta$ and $d_{\textrm{r}}$, respectively, from the AP. We assume that a far end user is horizontally distant from the AP by a distance $d_{\textrm{u}}$, and no direct VLC link exists between the AP and the end user.
Let $T_{\textrm{tot}}^{(i)}$ denote the~$i^{\textrm{th}}$ block transmission time, measured in seconds. Additionally, $\tau_{i}$~(unitless) represents the portion of time allocated to transmit information and energy to the relay node in the $i^{\textrm{th}}$ time block. Thus, the duration of this phase is $T_{\textrm{VLC},{i}} = \tau_{i} T_{\textrm{tot}}^{(i)}$ seconds. We assume that the block transmission time is constant; hence, we drop the superscript of $T_{\textrm{tot}}^{(i)}$ in the sequel to simplify notation. Fig.~\ref{fig:transmission_block} depicts the transmission block under consideration. Without loss of generality, we assume that $T_{\textrm{tot}} = 1$ second.

\subsection{VLC Link}\label{sub_vlc}
In the first hop, the LED transmits both energy and information to the relay node through the VLC link. To ensure the non-negativity of the transmitted optical signal, a DC bias, denoted by $I_{\textrm{b},i}$, is added to the modulated signal. Specifically, the transmitted optical signal is expressed as
${{x_{\textrm{t},{i}}}\left( t \right) = P_{\textrm{LED}}\big( {{x_{\textrm{s},{i}}}\left( t \right) + {I_{{\textrm{b}},{i}}}} \big)}$ where $P_{\textrm{LED}}$
where $P_{\textrm{LED}}$ represents the LED power per unit (in W/A) and $x_{\textrm{s},i}(t)$ is the modulated electrical signal. We assume that the information-bearing signal is zero-mean and satisfies the peak-intensity constraint of the optical channel, such that~\cite{1}
\begin{equation}\label{eq:1}
A_{{i}} \le \min \left( {{I_{\textrm{b},{i}}} - {I_{\textrm{min}}},{I_{\textrm{max}}} - {I_{\textrm{b},{i}}}} \right),
\end{equation}
where $A_{{i}}$ denotes the peak amplitude of the input electrical signal (i.e., $\max (\left| {{x_{\textrm{s},{i}}}(t)} \right|) = A_{{i}}$), and ${I_{\textrm{b},{i}}} \in \left[ {{I_{\textrm{min}}},{I_{\textrm{max}}}} \right]$ with ${I_{\textrm{max}}}$ and ${I_{\textrm{min}}}$ being the maximum and minimum input currents of the DC offset, respectively. 
Let ${B_{{\textrm{VLC}}}}$ denote the double-sided signal bandwidth. 

Then, the information rate associated with the optical link between the AP and relay node within a block with $T_{\textrm{tot}} = 1$ second, is given as~\cite{1}
\begin{equation}\label{eq:2}
{R_{{\textrm{VLC}},{i}}} = T_{\textrm{VLC},{i}}{B_{{\textrm{VLC}}}}{\log _2}\left( {1 + \frac{e}{{2\pi }}\frac{{{{\left( {\eta P_{\textrm{LED}}A_{{i}}{H_{\textrm{VLC}}}} \right)}^2}}}{{\sigma _{{\textrm{VLC}}}^2}}} \right),
\end{equation}
where $\eta$ is the photo-detector responsivity in A/W and ${H_{{\textrm{VLC}}}}$ is the optical DC channel gain. In \eqref{eq:2}, $\sigma _{{\textrm{VLC}}}^2$ is the power of shot noise at the PD which is given as $\sigma _{{\textrm{VLC}}}^2 = {q_{\textrm{e}}}{I_{i}}{B_{{\textrm{VLC}}}}$ where ${q_{\textrm{e}}}$ is the charge of an electron and ${I_{i}}$ is the induced current due to the ambient light. One should note that the shot noise is the dominant noise source in the VLC channel and we ignore the thermal noise in our paper~\cite{komine2003integrated}. The optical DC channel gain of the VLC link can be written as
\begin{equation}\label{eq:3}
{H_{{\textrm{VLC}}}} = \frac{{\left( {m + 1} \right){A_{\textrm{p}}}}}{{2\pi \left( {h_\Delta ^2 + d_{\textrm{r}}^2} \right)}}{\cos ^m}\left( {{\phi _{\textrm{r}}}} \right)\cos \left( {{\theta _{\textrm{r}}}} \right)\Pi \left( {\left| {{\theta _{\textrm{r}}}} \right|,\Theta } \right),
\end{equation}
where ${\phi _r}$ and ${\theta _r}$ are the respective angle of irradiance and incidence, respectively. The Lambertian order is $m = {{ - 1} \mathord{\left/{\vphantom {{ - 1} {{{\log }_2}\left( {\cos \left( \Theta  \right)} \right)}}} \right.
 \kern-\nulldelimiterspace} {{{\log }_2}\left( {\cos \left( \Phi  \right)} \right)}}$ where $\Phi $ is the half-power beamwidth of the LED, and ${A_{\textrm{p}}}$ and $\Theta $ are the detection area and  field-of-view (FoV) of the PD, respectively. The function $\Pi \left( {x,y} \right)$ is 1 whenever $x \le y$, and is 0 otherwise.

The harvested energy at this phase can be computed as~\cite{1}
\begin{equation}\label{eq:4}
{E_{\textrm{1},{i}}} = 0.75T_{\textrm{VLC},{i}}{I_{{\textrm{DC}},{i}}}{V_{\textrm{t}}}\ln \left( {1 + \frac{I_{{\textrm{DC}},{i}}}{{{I_0}}}} \right),
\end{equation}
where ${V_{\textrm{t}}}$ is the thermal voltage, ${I_0}$ is the dark saturation current, and ${I_{{\textrm{DC}},{i}}}$ is the DC part of the output current given as 
$ I_{{\textrm{DC}},{i}} = \eta {H_{{\textrm{VLC}}}}P_{\textrm{LED}}{I_{\textrm{b},{i}}}$.
In the time period $T_{\textrm{RF},{i}} = 1-T_{\textrm{VLC},{i}}$, the aim is to maximize the harvested energy while the relay transmits the information to the far user over the RF link. Thus, during the second phase, the LED eliminates the alternating current (AC) part and maximizes the DC bias, i.e., $A_{{i}} = 0$ and $I_{\textrm{b},{i}} =
I_{\textrm{max}}$. Mathematically speaking, the harvested energy during the second phase can be expressed as
\begin{equation}\label{eq:5}
{E_{2,{i}}} = 0.75T_{\textrm{RF},{i}}{I_{\textrm{DC, max}}}{V_{\textrm{t}}}\ln \!\left(\! {1 + \frac{I_{\textrm{DC, max}}}{I_0}}\! \right),
\end{equation}
where $I_{\textrm{DC, max}} = \eta {H_{{\textrm{VLC}}}}P_{\textrm{LED}}{I_{\textrm{max}}}$.

The total harvested energy at the relay that can be utilized for transmitting the decoded symbol to the far user through an RF link can be calculated~as
\begin{equation}\label{eq:6}
\begin{split}
E_{\textrm{h},{i}} = &{E_{1,{i}}} + {E_{2,{i-1}}}\\
 = & z\bigg(T_{\textrm{VLC},{i}}I_{\textrm{b},{i}}\ln{\Big(1+\frac{\eta H_{\textrm{VLC}} P_{\textrm{LED}} I_{\textrm{b},{i}}}{I_0}\Big)} \\ 
 & + T_{\textrm{RF},{i-1}}I_{\textrm{max}}\ln{\Big(1+\frac{\eta H_{\textrm{VLC}} P_{\textrm{LED}} I_{\textrm{max}}}{I_0}\Big)}\bigg),
\end{split}
\end{equation}
where $z=0.75 \eta H_{\textrm{VLC}} P_{\textrm{LED}} V_{\textrm{t}}$, ${E_{2,{i}}}$ represents the harvested energy during the RF transmission in the previous transmission block. In this paper, we assume that the initial harvested energy is 0 (i.e., ${E_{2, 0}} = 0$). 

Note that in our energy harvesting model, the fill factor (FF) is incorporated as a constant term (e.g., 0.75) in~\eqref{eq:4} and~\eqref{eq:5}. The FF is a critical parameter in photovoltaic systems, representing the ratio of the maximum achievable power to the product of the open-circuit voltage and short-circuit current. Typically, the FF ranges between 0.7 and 0.8~\cite{peng2021end, li2011solar}, reflecting the efficiency of the energy conversion process. By including the FF, our model accounts for non-idealities in photovoltaic energy conversion, thereby enhancing the realism and accuracy of the harvested energy estimates.

As it can be readily checked from~\eqref{eq:1}, increasing $I_{\textrm{b}, i}$ leads to a decrease in $A_{i}$ and, consequently, it decreases the information rate associated with the VLC link. On the other hand, decreasing $I_{\textrm{b}, i}$ limits the harvested energy that can be obtained during VLC transmission (i.e., $E_{\textrm{1},{i}}$). 

\subsection{RF Link}
In the second hop, the relay re-transmits the information to the far user through the RF link by utilizing the harvested energy. We assume that the energy used for data reception at the relay is practically negligible and the harvested energy is primarily employed for data transmission~\cite{rakia2016optimal, xiao2021cooperative}. The relaying operation is of DF type. Let ${B_{{\textrm{RF}}}}$ denote the bandwidth for the RF system and ${N_0}$ denote the noise power which can be defined as ${{N_0} = {P_0} + 10{\log _{10}}\left( {{B_{{\textrm{RF}}}}} \right) + {N_{\textrm{F}}}}$ where ${P_0}$ is the thermal noise power, and ${N_{\textrm{F}}}$ is the noise figure. Further, assume that the relay re-transmits the electrical signal with normalized power. The respective information rate is given as 
\begin{equation}\label{eq:7}
{R_{\textrm{RF},{i}}} = T_{\textrm{RF},{i}}{B_{{\textrm{RF}}}}{\log _2}\left( {1 + \frac{{{{P_{\textrm{h},{i}}}}{{\left| {{h_{{\textrm{RF}}}}} \right|}^2}}}{{{G_{{\textrm{RF}}}}{N_0}}}} \right),
\end{equation}
where ${h_{{\textrm{RF}}}}$ denotes the Rayleigh channel coefficients, ${{P_{\textrm{h},{i}}} = {E_{\textrm{h},{i}}}/T_{\textrm{RF},{i}}}$ is the transmit power and ${G_{{\textrm{RF}}}}$ is the path loss model for RF link and can be expressed as
\begin{equation}\label{eq:8}
{G_{{\textrm{RF}}}} = {\left( {\frac{{4\pi {d_0}}}{\lambda }} \right)^2}{\left( {\frac{d_{\textrm{u}}}{{{d_0}}}} \right)^\beta },
\end{equation}
where $\lambda $ is the used RF carrier wavelength, ${d_0} = 1$~m is the reference distance, and $\beta $ is the path loss exponent, which generally takes a value between [1.6, 1.8]~\cite{rappaport1996wireless}. 

The achievable information rate is limited by the smaller information rate between the VLC link and the RF link and can be expressed as~\cite{guo2021achievable}
\begin{equation}\label{eq:9}
    R_{\textrm{VLC-RF},{i}} = \min{(R_{\textrm{VLC},{i}}, R_{\textrm{RF},{i}})}.
\end{equation}

\begin{figure}[t]
\centering
\begin{subfigure}{\linewidth} 
\centering
\includegraphics[trim=0.2cm 0.1cm 0.2cm 0.3cm, clip,width=\linewidth]{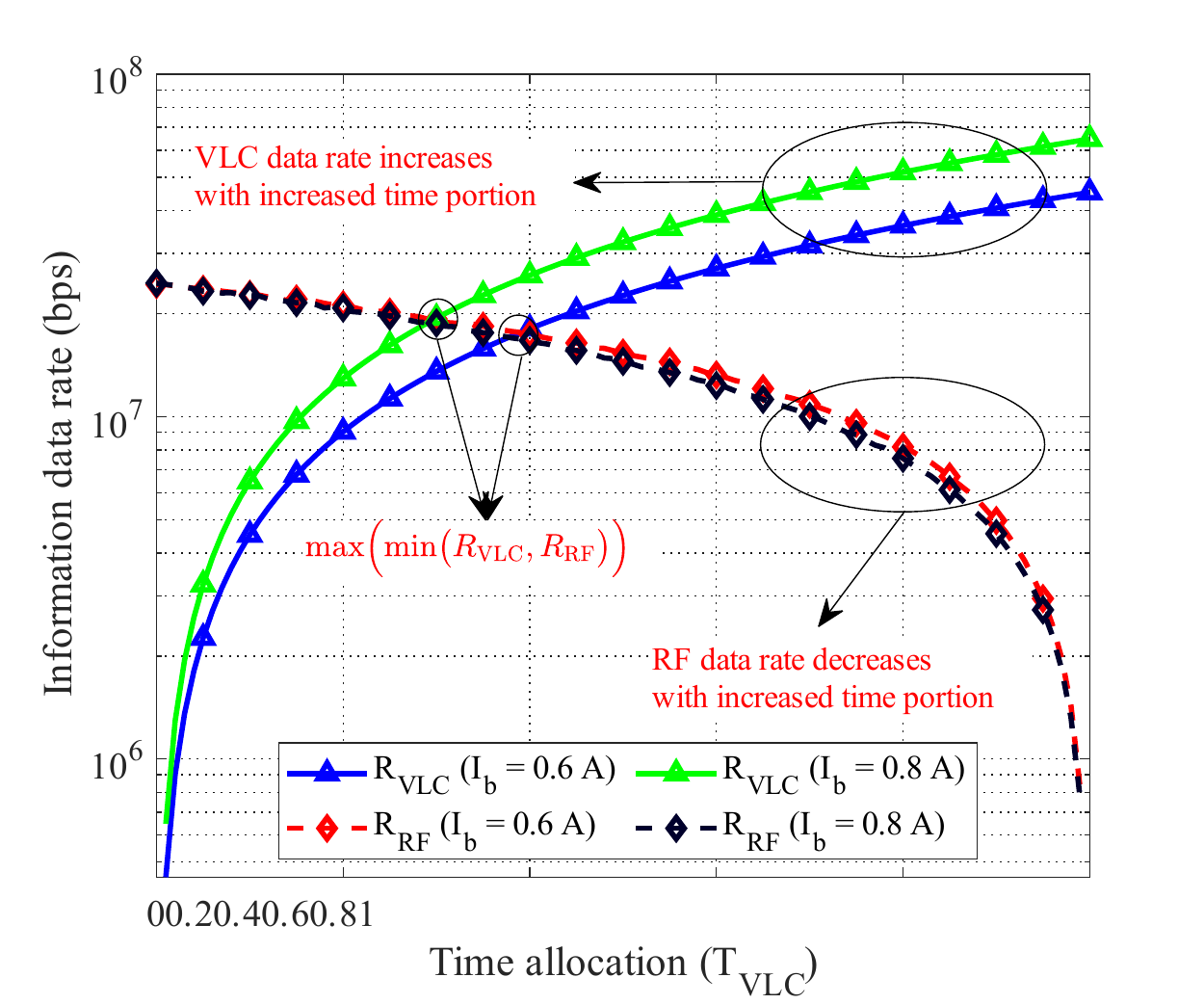}
\caption{Fixed $I_{\textrm{b}}$.}\label{fig:time_alc_efc} 
\end{subfigure}
\begin{subfigure}{\linewidth} 
\centering
\includegraphics[trim=0.2cm 0.1cm 0.2cm 0.3cm, clip,width=\linewidth]{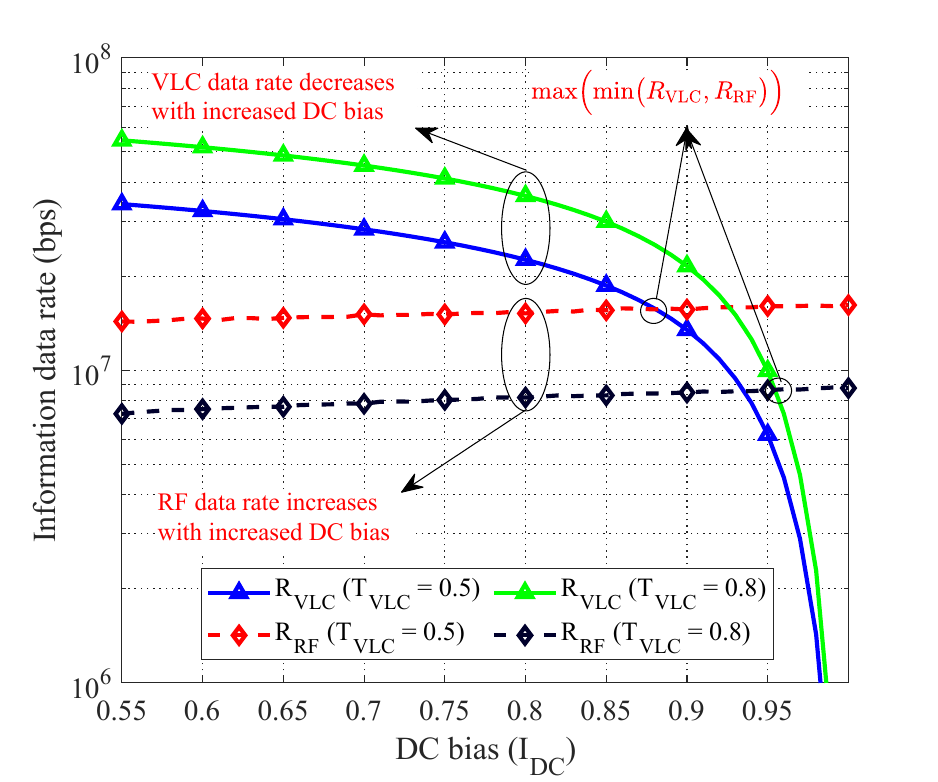} 
\caption{Fixed $T_{\textrm{VLC}}$.}
\label{fig:dc_bias_efc} 
\end{subfigure} 
\caption{The VLC and RF information rate when $d_{\textrm{r}} = 0$~m, $d_{\textrm{u}} = 4$~m, and $f_{\textrm{c}} = 2.4$~GHz for \textbf{(a)} fixed DC bias and \textbf{(b)} equal time allocation.}
\label{fig:effect_I_t}
\end{figure}

Fig.~\ref{fig:effect_I_t} illustrates the effect of time allocation and DC bias on information data rate. Unless otherwise stated, the system and channel parameters can be found in Table~\ref{table1}. We assume the relay is located at $d_{\textrm{r}} = 0$~m, the user is at $d_{\textrm{u}} = 4$~m, and the RF frequency sets as $f_{\textrm{c}} = 2.4$~GHz. 
Fig.~\ref{fig:time_alc_efc} illustrates the information data rate for the VLC and RF links versus the time allocation of the VLC link. In this figure, we assume DC bias as $I_{\textrm{b}} \in \{0.6, 0.8\}$~A. 
Since we assume that the block transmission time is constant ($T_{\textrm{VLC}, i} + T_{\textrm{RF}, i} = 1$), as the VLC time portion increases the RF time portion decreases. 
As it can be observed from Fig.~\ref{fig:time_alc_efc}, increasing the time allocation for the VLC link (i.e., $T_{\textrm{VLC}}$) results in increasing the VLC data rate while it decreases the harvested energy during the second phase (see~\eqref{eq:5}) and consequently decreases the RF data rate.
Fig.~\ref{fig:dc_bias_efc} depicts the information rate for VLC and RF link versus DC bias. Here, we assume an equal time portion for VLC and RF transmission (i.e., $T_{\textrm{VLC}} = T_{\textrm{RF}} = 0.5$) as well as $T_{\textrm{VLC}} = 0.8$ (consequently $T_{\textrm{RF}} = 0.2$). Recalling \eqref{eq:1}, we can observe that increasing DC bias leads to a reduction in peak amplitude of the input electrical signal (i.e., $A_i$) and subsequently VLC data rate. However, as DC bias increases the harvested energy in the first phase (see \eqref{eq:4}) increases which eventually results in a higher RF data rate.

\begin{figure*}[t]
\centering
\includegraphics[width=0.9\linewidth]{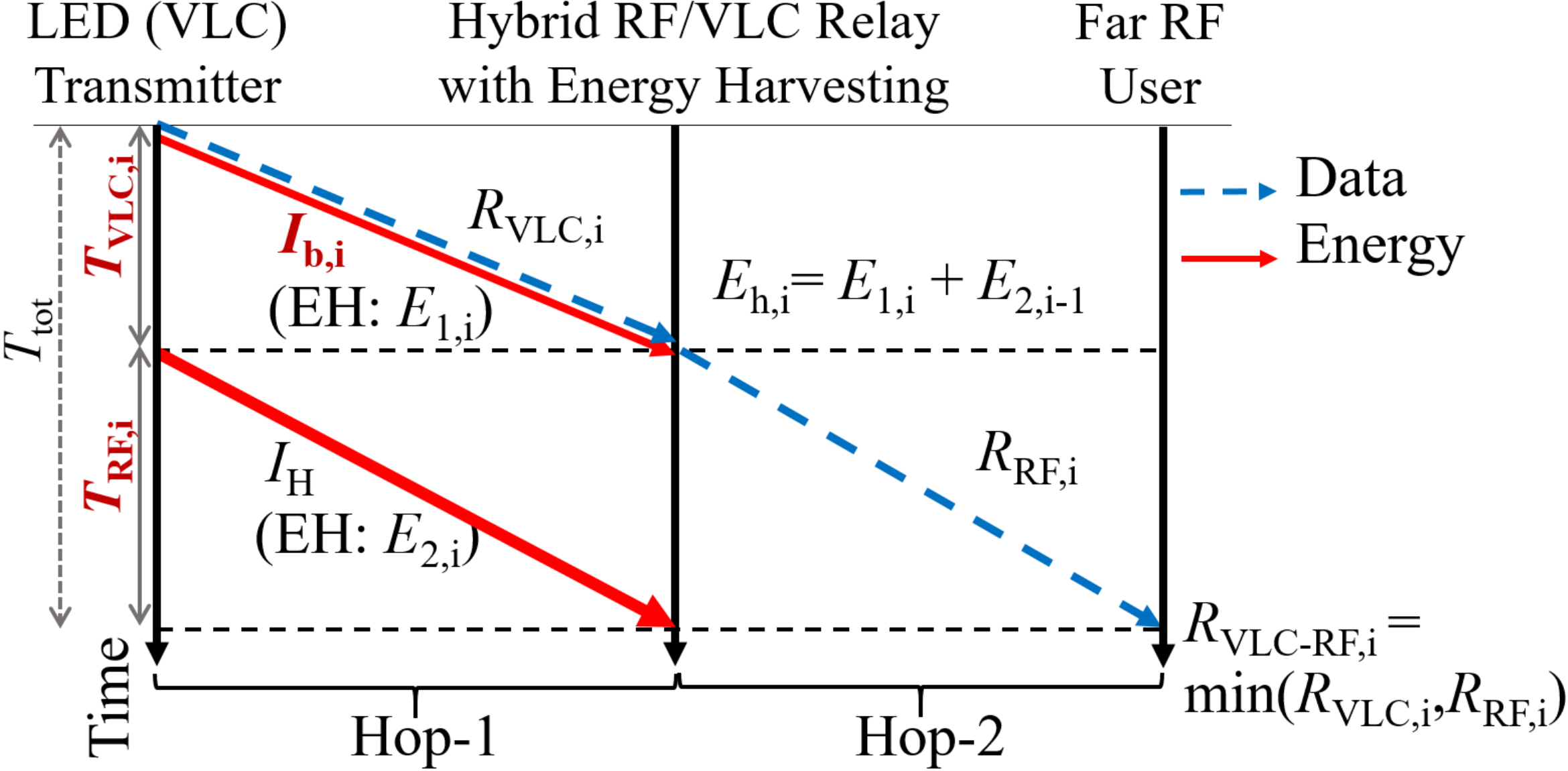}
\caption{Summary of the optimization problem that involves LED transmitter, hybrid RF/VLC relay with energy harvesting, and the far RF user that receives the data through the relay node through an RF link. Here, we consider $I_{\textrm{b}, i}$, $T_{\textrm{VLC}, i}$, and $T_{\textrm{RF}, i}$ as the optimization variables to maximize the end-to-end system data rate.}
\label{fig:schematic_sys}
\end{figure*}

\section{Optimization Framework}\label{sec:opt_fw}
Our aim is to optimize the achievable information rate (i.e.,~\eqref{eq:9}) over $I_{\textrm{b}, i}$ and $T_{\textrm{VLC}, i}$. Fig.~\ref{fig:schematic_sys} summarizes the optimization problem of maximizing the system data rate, with the optimization variables represented in red.
Recalling the information rate in VLC link (i.e., \eqref{eq:2}) and RF link (i.e., \eqref{eq:7}), the optimization problem can be written as
\begin{equation}\label{eq:10}
\begin{aligned}
\max_{I_{\textrm{b},{i}}, T_{\textrm{VLC},{i}}, T_{\textrm{RF},{i}}} \ & R_{\textrm{VLC-RF},{i}}\\
\textrm{s.t.} \quad & c_1: I_{\textrm{min}} \leq I_{\textrm{b},{i}} \leq I_{\textrm{max}}, \\
  & c_2:  T_{\textrm{VLC},{i}} \!+\! T_{\textrm{RF},{i}} \!=\! 1, T_{\textrm{VLC},{i}} >0, T_{\textrm{RF},{i}} > 0  \\
  & c_3:  R_{\textrm{th}} \leq R_{\textrm{RF}}^{i}~,
\end{aligned}
\end{equation}
where $R_{\textrm{th}}$ is a predefined threshold value, and constraint $c_1$ is imposed to avoid any clipping and guarantee that the LED operates in its linear region. Since the relay re-transmits the information and the RF far user is unable to receive data from the LED, $c_3$ is added to satisfy the minimum required data rate. 

The joint-optimization problem in \eqref{eq:10} is non-smooth (due to the $\min$ operator) and non-convex (due to the objective function and constraint $c_3$). We reformulate the above optimization problem in the epigraph form to remove the non-smoothness in the objective function. Referring to~\cite[Chapter~4]{boyd2004convex}, the epigraph form of~\eqref{eq:10} can be written as
\begin{equation}\label{eq:11}
\begin{aligned}
&\max_{\phi, I_{\textrm{b},{i}}, T_{\textrm{VLC},{i}}, T_{\textrm{RF},{i}}} \quad  \phi \\
\textrm{s.t.} \quad & c_1, c_2, c_3,\\
\quad  & c_4:  \phi \leq R_{\textrm{VLC},{i}} \\
\quad  & c_5: \phi \leq R_{\textrm{RF},{i}}.
\end{aligned}
\end{equation}
The above equivalent optimization problem to \eqref{eq:10} solves the non-smoothness, while it is still non-convex. Let ${\alpha = {e\left( {\eta P_{\textrm{LED}}{H_{\textrm{VLC}}}} \right)}^2/(2\pi\sigma _{{\textrm{VLC}}}^2)}$, ${\beta = \eta {H_{{\textrm{VLC}}}}P_{\textrm{LED}}}$, and
${\zeta = |h_{\textrm{RF}}|^2/(G_{\textrm{RF}}N_0)}$. Substituting~\eqref{eq:2} and~\eqref{eq:7} in~\eqref{eq:11}, we~have
\begin{equation}\label{eq:12}
\begin{aligned}
 &\max_{\phi, I_{\textrm{b},{i}}, A_{i}, E_{\textrm{h},{i}}, T_{\textrm{VLC},{i}},T_{\textrm{RF},{i}}} \  \phi \\
\textrm{s.t.} \quad & c_1^{\prime}, c_2, \\
\quad & c_3:  T_{\textrm{RF},{i}} B_{\textrm{RF}}\! \log_2\!\!\left(\!\!1 \!+\! \frac{\zeta E_{\textrm{h},{i}}}{T_{\textrm{RF},{i}}}\!\right)\! \!\geq \! \! R_{\textrm{th}},\\
\quad  & c_4: T_{\textrm{VLC},{i}} B_{\textrm{VLC}} \log_2(1 + \alpha A_{i}^2)\!\! \geq \! \phi ,\\
\quad  & c_5: T_{\textrm{RF},{i}} B_{\textrm{RF}} \log_2\!\left(\!1 \! + \! \frac{\zeta E_{\textrm{h},{i}}}{T_{\textrm{RF},{i}}}\right) \! \geq \! \phi ,\\
\quad  & c_6:   \min \big( I_{\textrm{b},{i}} - I_{\textrm{min}}, I_{\textrm{max}} - I_{\textrm{b},{i}}\big) \! \geq \! A_{i},\\
\quad  & c_7:  z\Big( T_{\textrm{VLC},{i}}I_{\textrm{b},{i}}\ln{\big(1 + \frac{\beta I_{\textrm{b},{i}}}{I_0}\big)} \\
\quad & \quad +\! T_{\textrm{RF},{i-1}} I_{\textrm{max}}\! \ln\!{\big( 1\! + \!\frac{\beta I_{\textrm{max}}}{I_0}\big)} \!\Big)\!\! \geq \! E_{\textrm{h},{i}}.
\end{aligned}
\end{equation}

In the optimization problem of \eqref{eq:12},  $A_{i}^2$ is used in $c_4$ and $c_6$ is still non-smooth. Here, we relax $c_6$ by using Proposition 1 from~\cite{guo2021achievable}. Intuitively, as $I_{\textrm{b,i}}$ increases the harvested energy increases; however, it has a negative effect on the rate beyond $(I_{\textrm{min}} + I_{\textrm{max}})/2$. Thus, the optimal value of the term $I_{\textrm{b,i}}$ would be within $(I_{\textrm{min}} + I_{\textrm{max}})/2$ and $I_{\textrm{max}}$ (and not the other regime ${0 \leq I_{\textrm{b, i}} \leq I_{\textrm{max}})}$. The above restriction enforces $0 \leq A_{i} \leq  I_{\textrm{max}} - I_{\textrm{b},{i}}$ benefiting in getting rid of the non-smooth $\min$ operator (see~\eqref{eq:12}) as well. This leads to ${c_1^{\prime}: (I_{\textrm{min}} + I_{\textrm{max}})/{2} \leq I_{\textrm{b}, i} \leq I_{\textrm{max}}}$ and $c_6^{\prime}: 0 \leq A_{i} \leq  I_{\textrm{max}} - I_{\textrm{b},{i}}$. The constraints $c_3, c_4, c_5$, and $c_7$ are jointly non-convex. In this regard, we split the joint optimization problem into two sub-problems and solve them in a cyclic fashion which will be elaborated in the next section.
\looseness = -1

\section{Solution Approach}\label{sec:sol_app}
In this section, we consider two sub-problems for solving \eqref{eq:12}. In sub-problem 1, we solve the maximization problem for $\phi$ over $I_{\textrm{b},{i}}, A_{i}, E_{\textrm{h},{i}}$ by fixing the time allocation  $T_{\textrm{VLC},{i}}$. In the second sub-problem, we solve the maximization problem for $\phi$ over  $T_{\textrm{VLC},{i}}, T_{\textrm{RF},{i}},  E_{\textrm{h},{i}}$ by using $I_{\textrm{b},{i}}$ obtained from sub-problem 1.

\subsection{Sub-problem 1} 
First, we fix $T_{\textrm{VLC},{i}}$ (and hence $T_{\textrm{RF},{i}} = 1- T_{\textrm{VLC},{i}}$)  and solve the maximization problem for $\phi$ over $I_{\textrm{b},{i}}, A_{i}, E_{\textrm{h},{i}}$. Sub-problem 1 can be written as
\begin{equation}\label{eq:13}
\begin{aligned}
\max_{\phi, I_{\textrm{b},{i}},A_{i},E_{\textrm{h},{i}} } \quad & \phi \\
\textrm{s.t.} \quad & c_1^{\prime}, c_3, c_4, c_5, c_7\\
\quad &  c_6^{\prime}:   0 \leq A_{i} \leq I_{\textrm{max}} - I_{\textrm{b},{i}},
\end{aligned}
\end{equation}
where the constraints $c_3, c_4, c_5$ are conditionally convex.

\textbf{Assumption:}
The typical illumination requirement in an indoor VLC environment results in a high transmit optical intensity, which can provide a high SNR at the receiver~\cite{hanzo2012wireless, wang2018physical}. In this paper, we assume that SNR for the VLC link is much greater than 1 (in linear scale); i.e., $\alpha (A_{i})^2 \gg 1$. In this condition, we further utilize ${\log(1+x) \approx \log(x)}$ in the constraints $c_4$. Thus, the optimization problem can be written~as
\begin{equation}\label{eq:14}
\begin{aligned}
\max_{\phi, I_{\textrm{b},{i}},A_{i},E_{\textrm{h},{i}}} \quad  & \phi \\
\textrm{s.t.} \quad & c_1^{\prime}, c_3, c_5, c_6^{\prime}\\
\quad & c_4^{\prime}:  T_{\textrm{VLC},{i}} B_{\textrm{VLC}} \log_2(\alpha A_{i}^2) \geq \phi ,\\
\quad  & c_7^{\prime}:  z\bigg( T_{\textrm{VLC},{i}}I_{\textrm{b},{i}}\ln{\Big(\frac{\beta I_{\textrm{b},{i}}}{I_0}\Big)} \\
 &\quad +  \! T_{\textrm{RF},{i-1}} I_{\textrm{max}}\ln\!{\Big(\! \frac{\beta I_{\textrm{max}}}{I_0}\Big)} \bigg)\! \geq \! E_{\textrm{h},{i}}.
\end{aligned}
\end{equation}
In \eqref{eq:14}, $c_4^{\prime}$ is a convex constraint while $c_7^{\prime}$ is still not convex. We further utilize the first-order Taylor series and MM approach to relax this constraint~\cite{Yusuf_VLC,sun2016majorization}. As a result, $c_7^{\prime}$ can be replaced with 
\begin{equation}\label{eq:15}
c_7^{\star}: \quad  g(I_{\textrm{b},{i}}) = g_0\big(I_{\textrm{b},{i}}(t)\big) + \frac{\partial g\big(I_{\textrm{b},{i}}(t)\big)}{\partial I_{\textrm{b},{i}}} \big(I_{\textrm{b},{i}} - I_{\textrm{b},{i}}(t)\big),
\end{equation}
where 
\begin{equation}\label{eq:16}
\begin{aligned}
g_0(I_{\textrm{b},{i}}(t)) =&  z\bigg(T_{\textrm{VLC},{i}} I_{\textrm{b},{i}}(t)\ln{\Big(\frac{\beta I_{\textrm{b},{i}}(t)}{I_0}\Big)}\\ & +
 T_{\textrm{RF},{i}-1}I_{\textrm{max}}\ln{\Big(\frac{\beta I_{\textrm{max}}}{I_0}\Big)}\bigg),
 \end{aligned}
 \end{equation}
and\\
 \begin{equation}\label{eq:17}
 \begin{aligned}
 \frac{\partial g(I_{\textrm{b},{i}}(t))}{\partial I_{\textrm{b},{i}}} & = z T_{\textrm{VLC},{i}}\bigg(\ln{\Big(\frac{\beta I_{\textrm{b},{i}}(t)}{I_0}\Big)} + \frac{\beta I_{\textrm{b},{i}}(t)}{I_0 + \beta I_{\textrm{b},{i}}(t)} \bigg).
 \end{aligned}
  \end{equation}
In \eqref{eq:15}, the term $t$ is an index-term and denotes the iteration index for the MM approach. The MM procedure on \eqref{eq:15} operates iteratively. We first solve the problem for some initial values of $I_{\textrm{b},{i}}(t)$. Then, we update the value of $I_{\textrm{b},{i}}(t)$ at each iteration until it remains the same for two consecutive iterations, or the change between two consecutive iterations is not appreciable.

Overall, the optimization sub-problem 1 is as follows:
\begin{equation}\label{eq:18}
\begin{aligned}
& \max_{\phi, I_{\textrm{b},{i}},A_{i},E_{\textrm{h},{i}} } \quad  \phi \\
\textrm{s.t.} \quad & c_1, c_3, c_4^{\prime}, c_5, c_6^{\prime}, c_7^{\star}.
\end{aligned}
\end{equation}
We iteratively solve the above sub-problem 1 until its convergence. Once the above sub-problem converges, we continue with sub-problem 2 which is elaborated in the~following.\looseness=-1

\subsection{Sub-problem 2}
In here, we fix $I_{\textrm{b},{i}}$ obtained from sub-problem 1 and solve the problem for maximizing $\phi$ over the variables $T_{\textrm{VLC},{i}}, T_{\textrm{RF},{i}},  E_{\textrm{h},{i}}$. The optimization problem can be expressed~as
\begin{equation}\label{eq:19}
\begin{aligned}
& \max_{\phi, T_{\textrm{VLC},{i}}, T_{\textrm{RF},{i}},E_{\textrm{h},{i}} } \quad  \phi \\
\textrm{s.t.} \quad & c_2,  c_3,  c_4^{\prime},  c_5, c_7^{\prime}.
\end{aligned}
\end{equation}
In \eqref{eq:19}, the objective function and constraints $c_2, c_4^{\prime}, c_7^{\prime}$ are linear, whereas the constraints $c_3$ and $c_5$ are convex constraints which result in a convex optimization problem.

Please note that the cyclic minimization framework helps reduce the number of non-convex constraints by decomposing the original problem into two subproblems (with independent variable) that are solved iteratively in a cyclic manner. Regarding theoretical guarantees, since the joint optimization problem is inherently non-convex, we cannot guarantee convergence to a global optimum; we can only ensure convergence to a stationary point.

\subsection{Convergence}
\begin{figure}[t!]
\centering
\includegraphics[trim=0.2cm 0.1cm 0.2cm 0.3cm, clip,width=\linewidth]{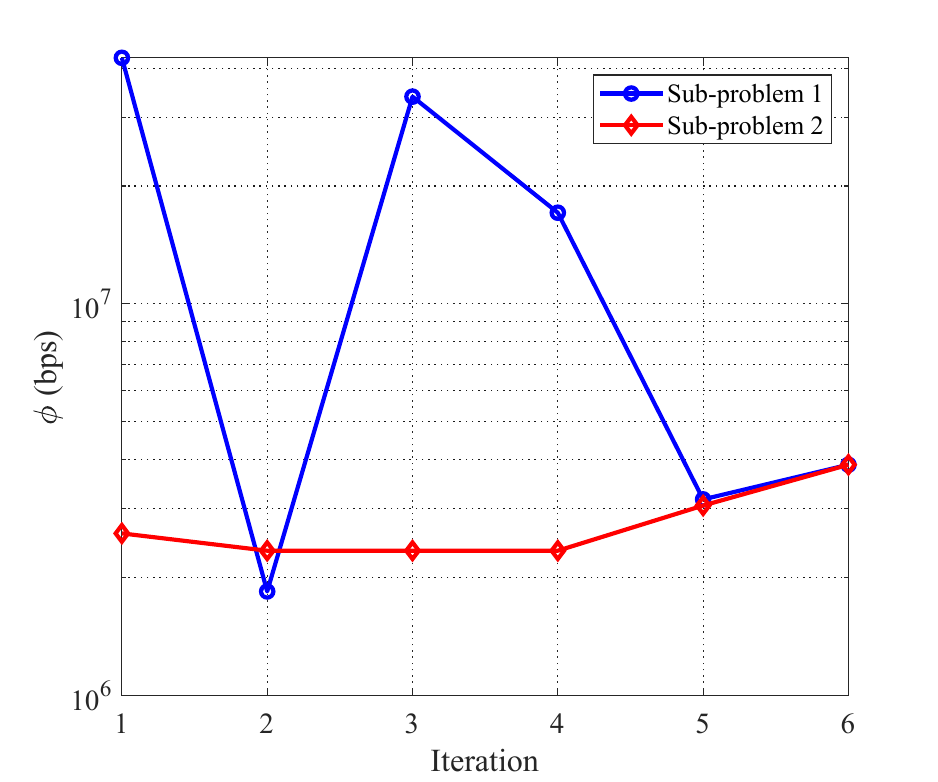}
\caption{The performance of our optimization framework versus the iteration count when $d_r = 0$~m and $d_u = 4$~m.}
\label{fig:conv}
\end{figure}

Here, we study the convergence of our proposed optimization algorithm. We assume that the relay location is at $d_r = 0$~m while the far user distance is $d_u = 4$~m. As it can be observed from Fig.~\ref{fig:conv}, the achievable information rate obtained for sub-problem 1 is higher than the one obtained for sub-problem 2.However, after five iterations, rates obtained in the two sub-problems converge and the difference between the information rate of the sub-problems becomes negligible.

\section{Relay Random Orientation}\label{sec:random_or}
In this section, we investigate the effect of relay random orientation on the achievable data rate for the considered scenario in Figs.~\ref{fig:system_model},~\ref{fig:transmission_block}, and \ref{fig:schematic_sys}. Random orientation can significantly influence channel quality. This effect not only degrades the VLC data rate but also influences the RF data rate since the relay is empowered by the harvested energy.
Here, we assume wide FoV where the incidence angle $\theta_{\textrm{r}}$ in \eqref{eq:3} is always smaller than $\Theta$ which implies $\Pi \left( {\left| \theta_{\textrm{r}}  \right|,\Theta } \right) = 1$; therefore the LED is always within the FoV. To separate the deterministic and random parts, we can rearrange \eqref{eq:3} as follows:
 \begin{equation}\label{eq:20}
 H_{{\textrm{VLC}}} = \frac{{\left( {m  + 1} \right){A_{\textrm{p}}}{h_\Delta^m }}}{{2\pi }}{\left( {{h_\Delta^2} + {d_{\textrm{r}}^2}} \right)^{ - \frac{{m  + 2}}{2}}}\cos \left( \theta_{\textrm{r}}  \right),
 \end{equation}
where we employ the geometrical relation $$\cos \left( {{\phi}_{\textrm{r}}} \right) =
\frac{h_\Delta}{\sqrt {d_{\textrm{r}}^2 + {h_\Delta^2}} }.$$
Let $H_{\textrm{VLC}} = h_{\textrm{c}}h_{\theta}$ where $${h_{\textrm{c}} = \frac{{\left( {m  + 1} \right){A_{\textrm{p}}}{h_\Delta^m }}}{{2\pi }}{\left( {{h_\Delta^2} + {d_{\textrm{r}}^2}} \right)^{ - \frac{{m  + 2}}{2}}}}$$ is the deterministic part of \eqref{eq:20} and $h_{\theta} = \cos(\theta_{\textrm{r}})$.  The distribution of the square channel can be derived by considering the probability density function (PDF) of $h_\theta^2=\cos^2(\theta_{\textrm{r}})$ given as~\cite{erouglu2018impact} \looseness=-1
\begin{equation} \label{eq:21}
    {f_{h_\theta ^2}}\left( x \right) = \frac{{{c_\theta }}}{{\sqrt {4x\left( {1 - x} \right)} }}{f_\theta }\left( {\frac{1}{2}{{\cos }^{ - 1}}\left( {2x - 1} \right)} \right),
    \end{equation}
for $0 \leq x \leq 1$, and 0 otherwise. In \eqref{eq:21}, $c_{\theta}$ is the normalization constant and $f_{\theta}(\cdot)$ is the PDF of the random angle $\theta$.
As a result, the PDF of the square channel is readily given as~\cite{erouglu2018impact}
\begin{equation} \label{eq:22}
    f_{h^2}(x) = \frac{1}{h_c^2}{f_{h_\theta ^2}}\Big(\frac{x}{h_c^2}\Big).
\end{equation}
\subsection{Average VLC data rate}
In this condition, the average data rate for the VLC link can be calculated as
\begin{equation}\label{eq:23}
\begin{split}
{R_{{\textrm{VLC}},{i}}}^{\textrm{avg}} = &\int T_{\textrm{VLC},{i}}{B_{{\textrm{VLC}}}}\\
& \times{\log _2}
 \left( {1 + \frac{e}{{2\pi }}\frac{{{{\left( {\eta P_{\textrm{LED}}A_{{i}}} \right)}^2}}}{{\sigma _{{\textrm{VLC}}}^2}}}x \right)f_{h^2}(x) \text{\textrm{d}}x,
\end{split}
\end{equation}
It is worth mentioning that $f_{\textrm{h}^2}$ is 0 for $x < 0$ and $x>h_{\textrm{c}}^2$. In this paper, we assume $\theta$ follows $\theta \sim U[\theta_1, \theta_2]$. As a result, We can rewrite \eqref{eq:22} as
\begin{equation}\label{eq:24}
    f_{h^2}(x) = \frac{{c_\theta }}{(\theta_2 - \theta_1){\sqrt {4x\left( {1 - x} \right)} }}.
\end{equation}
Inserting \eqref{eq:24} into \eqref{eq:23}, and substituting $t \coloneqq x/h_{\textrm{c}}^2$ we have
\begin{equation}\label{eq:25}
{R_{{\textrm{VLC}},{i}}}^{\textrm{avg}} = L_1\int
\frac{{\log _2}{ \left( {1 + L_2}t \right)}}{{\sqrt{4t\left( {1 - t} \right)} }} \text{\textrm{d}}t,
\end{equation}
where $L_1 = \frac{{c_\theta }T_{\textrm{VLC},{i}}{B_{{\textrm{VLC}}}}}{\theta_2 - \theta_1}$ and $L_2 =\frac{e}{{2\pi }}\frac{{{{\left( {\eta P_{\textrm{LED}}A_{{i}}} {h_{\textrm{c}}}\right)}^2}}}{{\sigma _{{\textrm{VLC}}}^2}}$.
Utilizing the first two terms of Puiseux series\cite{winkler2016commutative} $x = 1$ for ${1 \mathord{\left/
 {\vphantom {1 {\sqrt {4x(1 - x)} }}} \right.
 \kern-\nulldelimiterspace} {\sqrt {4x(1 - x)} }}$, a lower bound on~\eqref{eq:25} can be written as
\begin{equation}\label{eq:26}
\begin{split}
{R_{{\textrm{VLC}},{i}}}^{\textrm{avg}}\! \geq \!  L_1 \! \Bigg(\!\!\int\!\!
\frac{{\log _2}{\left( {1\! + \! L_2}t \right)}}{2{\sqrt{ \left( {1 - t} \right)}} } \text{\textrm{d}}t  
\!\! + \!\!\int\!\!
\frac{\sqrt{1-t} {\log _2} \left( {1\! + \! L_2}t \right) }{4} \text{\textrm{d}}t \!\!\Bigg).
\end{split}
\end{equation}
Using 2.727.5 of~\cite{gradshteyn2014table}, the first integral term of \eqref{eq:26} can be written~as
\begin{equation}\label{eq:27}
\begin{split}
f_{1}(x, L_2) &=   \int
\frac{{\log _2} \left( {1 + L_2}t \right)}{2{\sqrt{\left( {1 - t} \right)}} } \text{\textrm{d}}t  \\
& = \frac{-1}{\sqrt{L_2}}\bigg( \left(\ln(x)-2\right) \sqrt{L_2+1-x} \\
& - 2\sqrt{L_2+1}\ln{\frac{\sqrt{L_2+1-x}-\sqrt{L_2+1} } {\sqrt{x}}} \bigg).
\end{split}
\end{equation}
Using an integral solver~\cite{wolframalpha}, the second integral in \eqref{eq:26} can be written as
\begin{equation}\label{eq:28}
\begin{split}
& f_{2}(x, L_2) =  \int
\frac{\sqrt{1-t} {\log _2} \left( {1 + L_2}t \right) }{4} \text{\textrm{d}}t \\
& = \frac{1}{4}\! \Bigg(\!\frac{2\sqrt{1\!-\!x}(-2L_2(x\!-\!4)\!\! +\!\! 3L_2(x\!-\!1)\ln{(L_2 x\!+\!1\!)}\! \!+\! \!6)}{9L_2} \\
&-\frac{4}{3}\left(\frac{L_2+1}{L_2}\right)^{(3/2)}{\tanh}^{-1}{\bigg(\sqrt{\frac{L_2(1-x)}{L_2+1}}\bigg)}\Bigg).
\end{split}
\end{equation}

The final result for~\eqref{eq:26} can then be expressed as
\begin{equation}\label{eq:29}
\begin{split}
{R_{{\textrm{VLC}},{i}}}^{\textrm{avg}}& \geq L_1 \\
& \times \bigg(f_{1}({\frac{\cos{(2 \theta _1)}+1}{2}}, L_2)+f_{2}(\!{\frac{\cos{(2 \theta_1)}+1}{2}}, L_2\!) \\
& -f_{1}({\frac{\cos{(2 \theta _2)}+1}{2}}, L_2)-f_{2}({\frac{\cos{(2 \theta _2)}+1}{2}}, L_2)\bigg).
\end{split}
\end{equation}

\subsection{Average Energy Harvesting}
Recalling \eqref{eq:6} and utilizing $\ln{(1+x)} \approx \ln{(x)}$, the total harvesting energy can be written as
\begin{equation}\label{eq:30}
\begin{split}
\tilde{E}_{\textrm{h},{i}} \approx M_1 H_{\textrm{VLC}}\ln{(M_2 H_{\textrm{VLC}})} + M_3 H_{\textrm{VLC}}\ln{(M_4 H_{\textrm{VLC}})} 
\end{split}
\end{equation}
where 
\begin{equation}
M_1 = 0.75T_{\textrm{VLC}}\eta P_{\textrm{LED}} I_{\textrm{b, i}} V_{\textrm{t}},    
\end{equation}
\begin{equation}
M_2 = \eta P_{\textrm{LED}} I_{\textrm{b, i}}/I_0,
\end{equation}
\begin{equation}
M_3 = 0.75T_{\textrm{RF}}\eta P_{\textrm{LED}} I_{\textrm{max}} V_{\textrm{t}},    
\end{equation}
and 
\begin{equation}
M_4 = \eta P_{\textrm{LED}} I_{\textrm{max}}/I_0.    
\end{equation}
The PDF of $h_{\theta}$ can be expressed as 
\begin{equation}
 f_{h_{\theta}}(x) = 1/(\theta_2 - \theta_1)\sqrt{(1-\cos^{-1}(x))}   
\end{equation}
for ${\cos{\theta_1}\leq x \leq \cos{\theta_2}}$. As a result, the PDF of VLC channel can be calculated as $f_{h}(x) = \frac{1}{h_c}f_{h_{\theta}}(\frac{x}{h_c})$. The average energy harvesting can be calculated as
\begin{equation}\label{eq:31}
\begin{split}
\Bar{E}_{\textrm{h},{i}} = & \int_0^\infty (M_1 x\ln{(M_2 x}) + M_3 x\ln{(M_4 x)} ) f_{h}(x)\text{\textrm{d}}x
\end{split}
\end{equation}
Utilizing $\ln{(ab)} = \ln{(a)} + \ln{(b)}$ and defining $f_3(x)$ and $f_4(x)$ as
\begin{equation}\label{eq:32}
\begin{split}
f_3(x) = & \int_{h_c\cos{\theta_2}}^{h_c\cos{\theta_1}} \frac{x}{\sqrt{1-\frac{x}{h_c}}}\text{\textrm{d}}x \\
= & -h_c\sqrt{h_c^2 - x^2}
\end{split}
\end{equation}
and
\begin{equation}\label{eq:33}
\begin{split}
&f_4(x) = \int_{h_c\cos{\theta_2}}^{h_c\cos{\theta_1}} \frac{x\ln{x}}{\sqrt{1-\frac{x}{h_c}}}\text{\textrm{d}}x \\
= &\! -\!h_c\Big(h_c\tanh^{-1}\big(\!\frac{\sqrt{h_c^2\!-\!x^2}}{h_c}\!\big)\!\! + \!\!\sqrt{h_c^2-x^2}\big(\ln{(x)}\!-\!1\big)\Big).
\end{split}
\end{equation}
Thus, the final expression for \eqref{eq:31} can be written as
\begin{equation}\label{eq:34}
\begin{split}
\Bar{E}_{\textrm{h},{i}} = & \frac{M_1}{h_c(\theta_2-\theta_1)}\bigg( \ln{(M_2)}f_3(h_c\cos{\theta_1}) + f_4(h_c\cos{\theta_1}) \\
&- \ln{(M_2)}f_3(h_c\cos{\theta_2}) - f_4(h_c\cos{\theta_2}) \bigg) \\
+ & \frac{M_3}{h_c(\theta_2-\theta_1)}\bigg( \ln{(M_4)}f_3(h_c\cos{\theta_1}) + f_4(h_c\cos{\theta_1}) \\
&- \ln{(M_4)}f_3(h_c\cos{\theta_2}) - f_4(h_c\cos{\theta_2}) \bigg).
\end{split}
\end{equation}
Therefore, a lower bound on the average data rate for RF link can be calculated~as
\begin{equation}\label{eq:35}
{R_{\textrm{RF},{i}}}^{\textrm{avg}} \geq T_{\textrm{RF},{i}}{B_{{\textrm{RF}}}}{\log _2}\left( {1 + \frac{\Bar{E}_{\textrm{h},i}{{{\left| {{h_{{\textrm{RF}}}}} \right|}^2}}}{T_{\textrm{RF}, i}{{G_{{\textrm{RF}}}}{N_0}}}} \right).
\end{equation}

\begin{table}[t]
\caption{System and channel parameters that are used to generate the numerical results.}
\label{table1}
\begin{center}
\scalebox{0.95}{
\begin{tabular}{ |l|l| } 
 \hline
 \textbf{Parameter} & \textbf{Numerical Value} \\ \hline
User distance ($d_{\textrm{u},{\textrm{min}}}$,$d_{\textrm{u},{\textrm{max}}}$) &  [4,8] m  \\ \hline 
Relay distance ($d_{\textrm{r},{\textrm{min}}}$,$d_{\textrm{r},{\textrm{max}}}$) & [0,2] m  \\ \hline 
LED power ($P_{\textrm{LED}}$) & 1.5 W/A \\ \hline
Noise figure ($N_{\textrm{F}}$) & 9 dB ~\cite{yapici2020energy} \\ \hline 
RF signal bandwidth $B_{\textrm{RF}}$ &  10 MHz~\cite{guo2021achievable} \\ \hline 
VLC signal bandwidth $B_{\textrm{VLC}}$ &  10 MHz~\cite{guo2021achievable} \\ \hline 
Thermal noise ($P_0$) & -174 dBm/Hz~\cite{yapici2020energy} \\ \hline 
RF frequency ($f_{\textrm{c}}$) &  \{2.4, 5\} GHz~\cite{yapici2020energy} \\ \hline 
Minimum DC bias ($I_{\textrm{min}}$) & 100 mA ~\cite{yapici2020energy}\\ \hline
Maximum DC bias ($I_{\textrm{max}}$) & 1 A ~\cite{yapici2020energy} \\ \hline
Photo-detector responsivity ($\eta$ ) & 0.4 A/W~\cite{1}\\ \hline
Thermal voltage ($V_{\textrm{t}}$) & 25 mV~\cite{yapici2020energy} \\ \hline
Dark saturation current ($I_0$) & $10^{-10}$ A~\cite{yapici2020energy} \\ \hline
FoV ($\Theta$) & $60^{\circ}$~\cite{yapici2020energy} \\ \hline
Half-power beamwidth ($\Phi$) & $60^{\circ}$~\cite{yapici2020energy} \\ \hline
Electron charge ($q_{\textrm{e}}$) & $1.6 \times 10^{-19}$ \\ \hline
Induced current ($I_{i}$) & $5840 \times10^{-6}$~\cite{yapici2020energy} \\ \hline
PD detection area ($A_{\textrm{p}}$) & $10^{-4}$ $\textrm{m}^2$~\cite{yapici2020energy} \\ \hline
AP relative height ($h_{\Delta}$) & 2 m~\cite{yapici2020energy} \\ \hline
Data rate threshold ($R_{\textrm{th}}$) & $10^6$ b/s \\ \hline
\end{tabular}}
\end{center}
\end{table}

\section{Numerical Results}\label{sec:num_results}
In this section, we evaluate the performance of the hybrid VLC-RF scheme depicted in Fig.~\ref{fig:system_model} using computer simulations. For the convenience of the reader, unless otherwise stated, the channel and system parameters are summarized in Table~\ref{table1}.

\subsection{Approximation and Closed-form Expressions}\label{sec:num_results_approx}

\begin{figure}[t!]
\centering
\includegraphics[trim=0.2cm 0.1cm 0.2cm 0.3cm, clip,width=\linewidth]{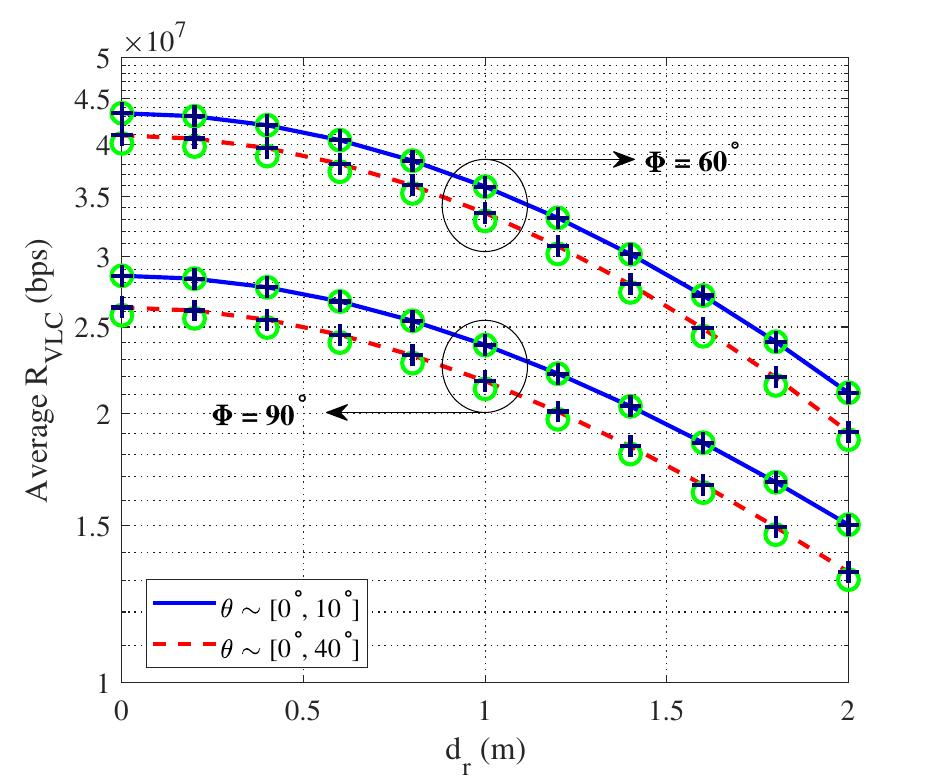}
\caption{The performance of VLC data rate through the exact expression (i.e., \eqref{eq:23}), simulation as illustrated with a black plus sign marker (\textbf{+}) and the closed-form (i.e., \eqref{eq:29}) as illustrated with a green circle marker (\textcolor{green}{\textbf{o}}) versus the horizontal distance between the relay and the VLC AP.}
\label{fig:average_VLC_random_orientation}
\end{figure}

\begin{figure}[!t]
\centering
\begin{subfigure}{1\columnwidth} 
\centering
\includegraphics[width=\textwidth]{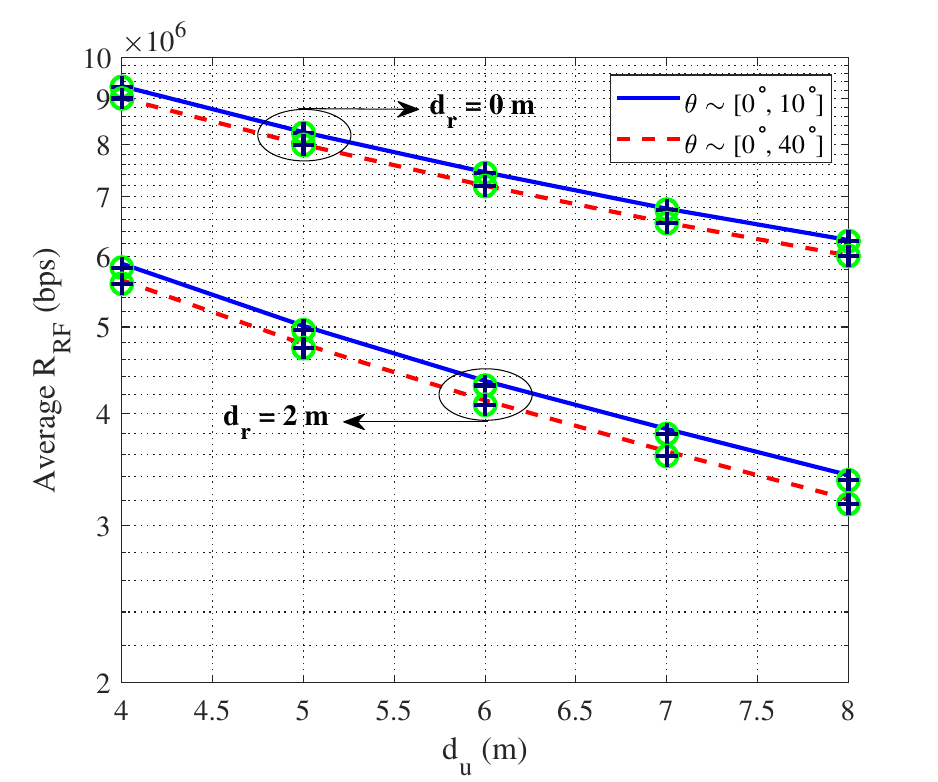} 
\caption{$\Phi = 60^{\circ}$.}\label{fig:average_RF_random_orientation_phi60} 
\end{subfigure}
\begin{subfigure}{1\columnwidth} 
\centering
\includegraphics[width=\textwidth]{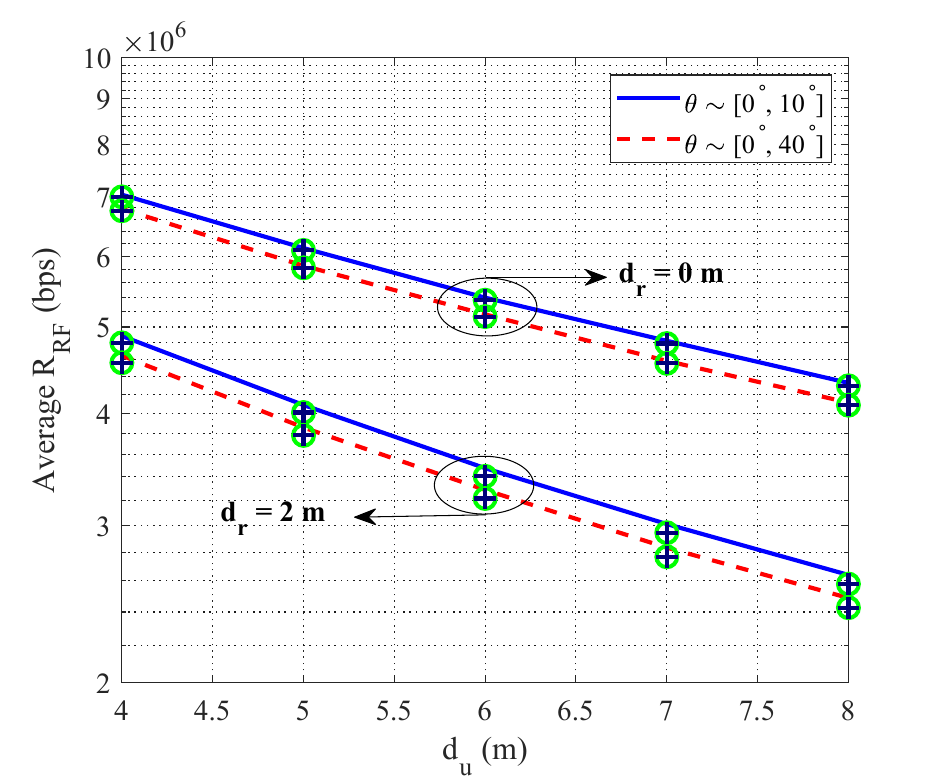}
\caption{$\Phi = 90^{\circ}$.}
\label{fig:average_RF_random_orientation_phi90}
\end{subfigure}
        \caption{The performance of RF data rate obtained using computer simulations, using \eqref{eq:31} as illustrated with circle marker and using the closed-form expression (i.e., \eqref{eq:34}) as illustrated with plus sign marker.}
        \label{fig:average_RF_random_orientation}
\end{figure}
In this subsection, we first compare the performance of VLC with random receiver orientation data rate using the exact expression in~\eqref{eq:23}, simulation, and the closed-form approximation in~\eqref{eq:29}. We consider two cases for the relay distance, where the relay location varies as $d_{\textrm{r}} \in [0, 2]$~m, and the half-power beamwidth values are $\Phi \in \{60^{\circ}, 90^{\circ}\}$. To validate our expressions, we analyze two cases of random receiver orientation: $\theta \sim U[0^{\circ}, 10^{\circ}]$ and $\theta \sim U[10^{\circ}, 40^{\circ}]$. 
We assume $A_{i}$ and $T_{\textrm{VLC}, i}$ remain constant throughout the communication, with $A = 0.2$ and $T_{\textrm{VLC}} = 0.8$.

Fig.~\ref{fig:average_VLC_random_orientation} illustrates the effect of random orientation on the average VLC data rate. As depicted in the figure, the simulation results perfectly align with the exact expression, confirming the accuracy of the derived formula. However, there is a minor, practically negligible discrepancy observed between the closed-form approximation and the exact solution.
From Fig.~\ref{fig:average_VLC_random_orientation}, it is evident that as the half-power beamwidth ($\Phi$) increases, the VLC data rate decreases. As indicated by~\eqref{eq:3}, an increase in the half-power beamwidth reduces the corresponding Lambertian order (i.e., $m$), which in turn decreases the optical DC channel gain for the on-axis receiver, and ultimately leads to a lower VLC data rate.
Notably, the reduction in data rate is more pronounced for larger orientation angles, as shown in the comparison between the two cases: $\theta \sim U[0^{\circ}, 10^{\circ}]$ and $\theta \sim U[10^{\circ}, 40^{\circ}]$, where the larger orientation range ($[10^{\circ}, 40^{\circ}]$) leads to a more significant decrease in the average data rate.

In addition to the impact of random receiver orientation on the VLC data rate, random orientation also influences the amount of harvested energy, which subsequently affects the data rate in the RF link. In Fig.~\ref{fig:average_RF_random_orientation}, we present the average RF data rate performance for two relay locations, specifically $d_{\textrm{r}} \in \{0, 2\}$~m, versus the RF user distance (i.e., $d_{\textrm{u}}$), while considering half-power beamwidths of $\Phi = 60^{\circ}$ and $\Phi = 90^{\circ}$. 
From Fig.~\ref{fig:average_RF_random_orientation}, it is evident that random receiver orientation affects the RF data rate, especially at greater relay distances. Additionally, as the half-power beamwidth increases from $\Phi = 60^{\circ}$ to $\Phi = 90^{\circ}$, the average RF data rate generally decreases, further corroborating the detrimental effect of larger beamwidths on energy harvesting efficiency and overall RF link performance. It is worth noting that this reduction is more evident when the relay is located closer to the transmitter (i.e., $d_{\textrm{r}}$ = 0~m). We will elaborate on this observation later in Section~\ref{sec:num_results_ro}.

Furthermore, the simulation results show close agreement with the derived closed-form expression for the system under consideration, particularly in scenarios with shorter relay node distances. This validates the accuracy of the closed-form approximation even when random receiver orientation is factored into the analysis.

\subsection{Achievable Data Rate without Relay Random Orientation}\label{sec:num_results_without_ro}
In this subsection, we analyze the achievable data rate for the system under consideration when the relay orientation is fixed and upward. We explore four distinct cases:

\begin{itemize} 

\item \textbf{Case 1:} Joint optimization (JO) of $I_{\textrm{b}}$ and $T_{\textrm{VLC}}$ while utilizing the harvested energy from the previous transmission block;

\item \textbf{Case 2:} JO of $I_{\textrm{b}}$ and $T_{\textrm{VLC}}$ without utilizing the harvested energy from the previous transmission block (i.e., $E_{2,{i-1}}=0$);

\item \textbf{Case 3:} Optimization of $I_{\textrm{b}}$ with fixed time allocation (FTA), where $T_{\textrm{VLC}}=T_{\textrm{RF}}=0.5$, utilizing the harvested energy from the previous transmission block (as in~\cite{rakia2016optimal});

\item \textbf{Case 4:} Optimization of $I_{\textrm{b}}$ with FTA, where $T_{\textrm{VLC}}=T_{\textrm{RF}}=0.5$ and without utilizing the harvested energy from the previous transmission block (i.e., $E_{2,{i-1}}=0$), similar to~\cite{peng2021end}.
\end{itemize}

\begin{figure}[t!]
\centering
\includegraphics[trim=0.2cm 0.1cm 0.2cm 0.3cm, clip,width=\linewidth]{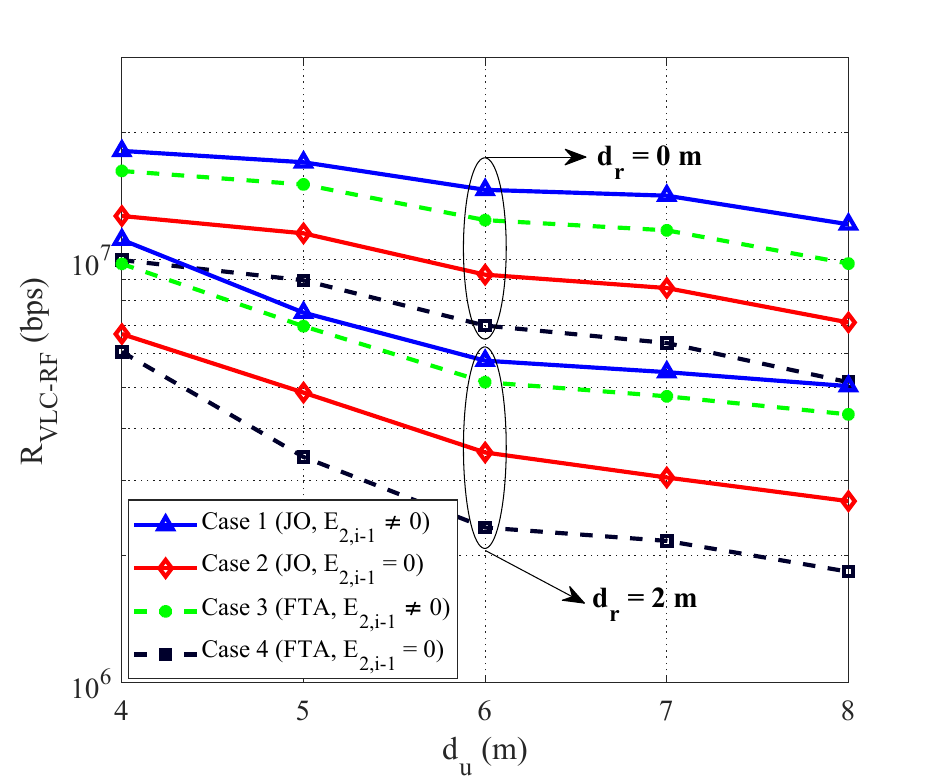}
\caption{The performance of optimal data rate for different user distances when the relay is located at $d_{\textrm{r}} = 0$~m.}
\label{fig:rate_diff_user_dis}

\end{figure}

Fig.~\ref{fig:rate_diff_user_dis} illustrates the optimal data rate for these four cases when the relay is positioned at $d_{\textrm{r}} = 0$~m and $d_{\textrm{r}} = 2$~m. The RF user distance varies between $d_{\textrm{u}} \in \{4, 5, 6, 7, 8\}$~m, and we assume an RF carrier frequency of $f_{\textrm{c}} = 2.4$~GHz.
The results depicted in Fig.~\ref{fig:rate_diff_user_dis} show that Cases 1 and 3, where the relay can harvest energy during RF transmission, significantly outperform Cases 2 and 4. This is due to the additional energy harvested, which leads to higher RF transmit power and supports higher data rates, particularly in energy-limited scenarios. As the user distance increases, a general decrease in the data rate is observed for all cases. This decline is primarily attributed to the increased path loss in the RF link.

\begin{figure*}[!ht]
\centering
\begin{subfigure}{0.32\linewidth} 
\centering
\includegraphics[trim=0.2cm 0.5cm 0.5cm 0.5cm, clip,width=\linewidth]{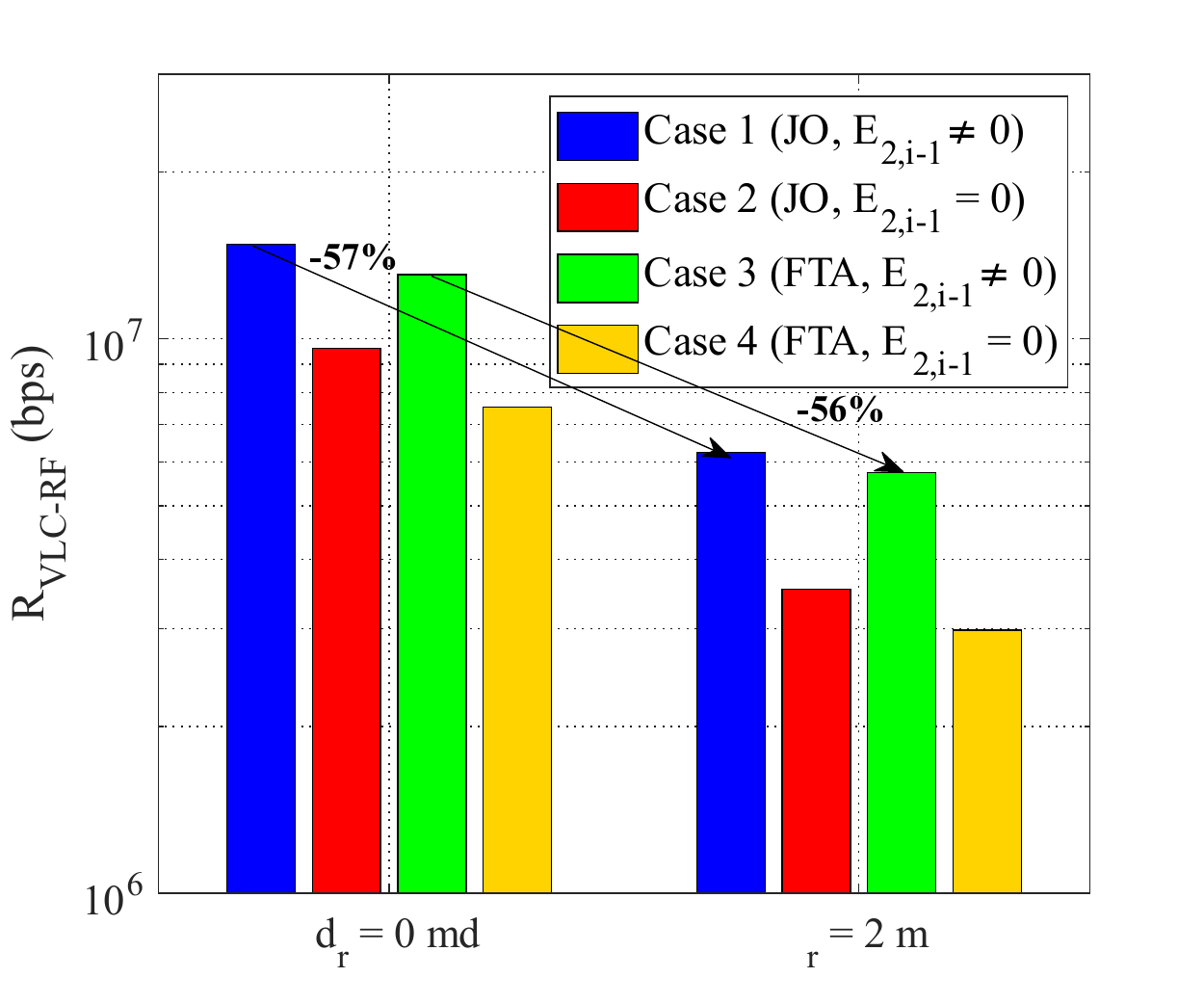} 
\caption{Optimal data rate}\label{fig:rate_random_2G} 
\end{subfigure}
\begin{subfigure}{0.32\linewidth} 
\centering
\includegraphics[trim=0.2cm 0.5cm 0.5cm 0.5cm, clip,width=\linewidth]{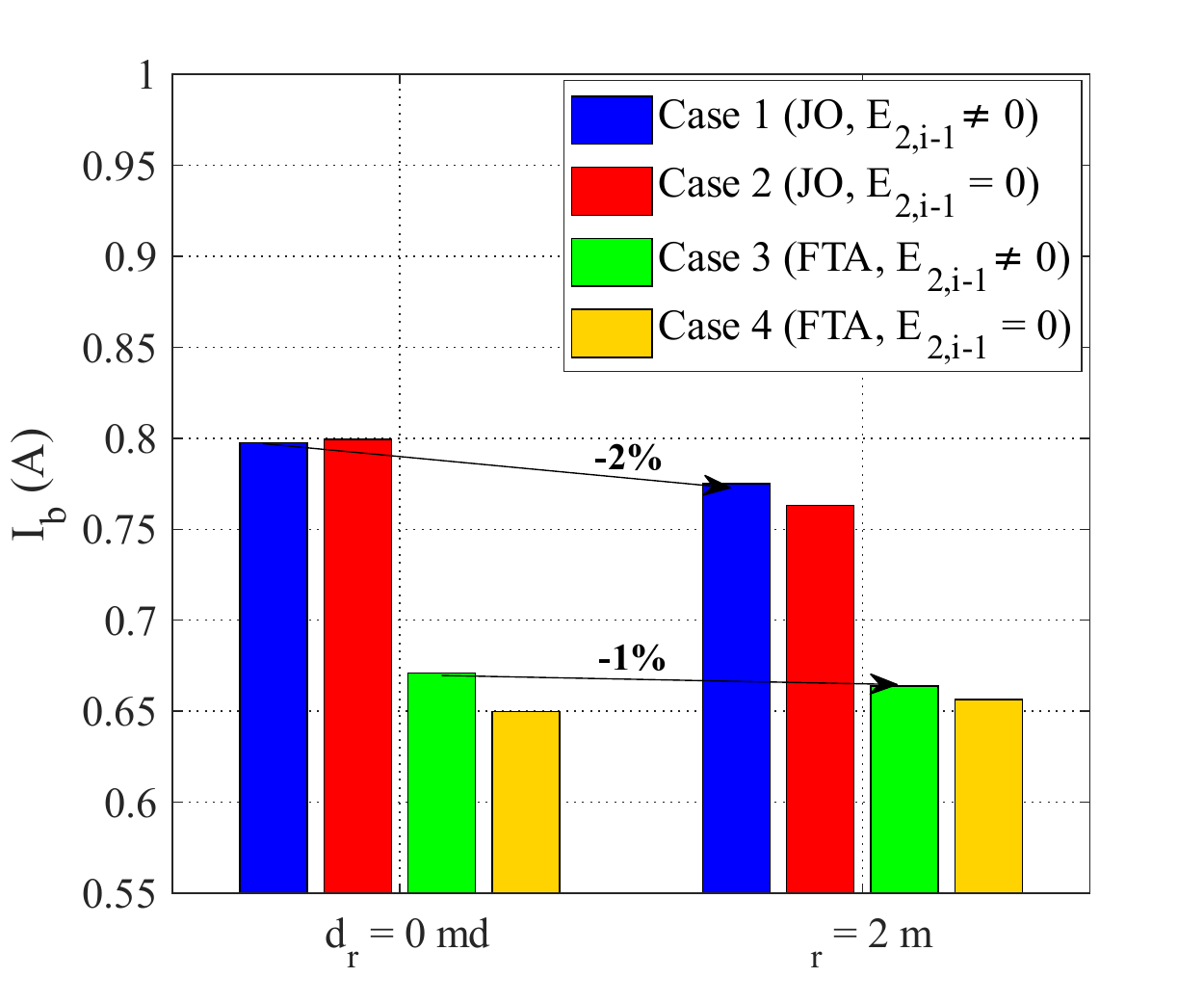} 
\caption{Optimal DC bias}
\label{fig:Ib_random_2G} 
\end{subfigure}
\begin{subfigure}{0.32\linewidth} 
\centering
\includegraphics[trim=0.2cm 0.5cm 0.5cm 0.5cm, clip,width=\linewidth]{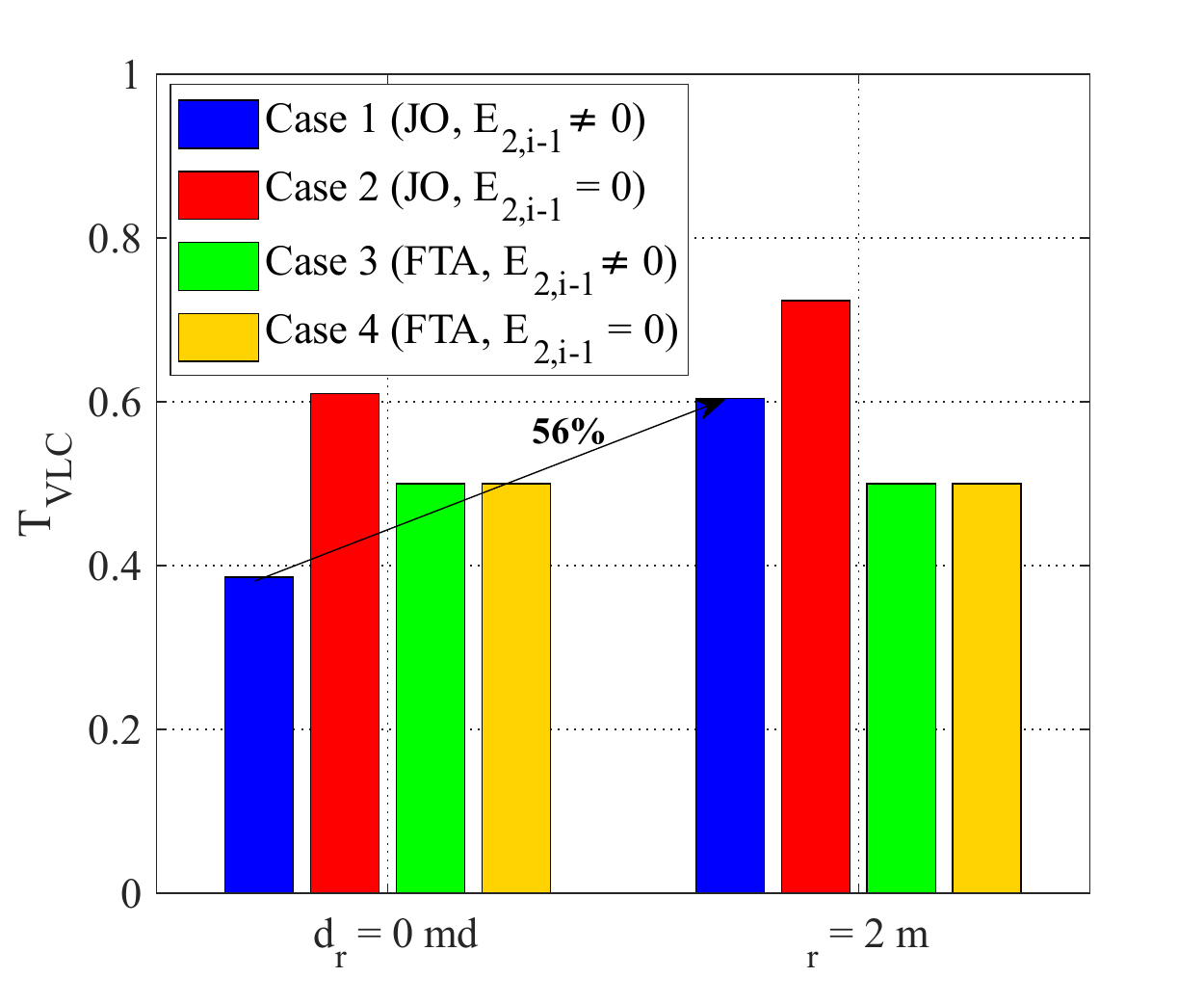} 
\caption{Optimal duration of VLC transmission}\label{fig:T_random_2G}
\end{subfigure} 
\caption{The performance of system under consideration with $f_{\textrm{c}} = 2.4$ GHz when the user node distance follows ${d_{\textrm{u}}} \sim \mathcal{U}\left[ {4, 8} \right]$.}
\label{fig:random_locations2G}
\end{figure*}

\begin{figure*}[t!]
\centering
\begin{subfigure}{0.32\linewidth} 
\centering
\includegraphics[trim=0.2cm 0.5cm 0.5cm 0.5cm, clip,width=\linewidth]{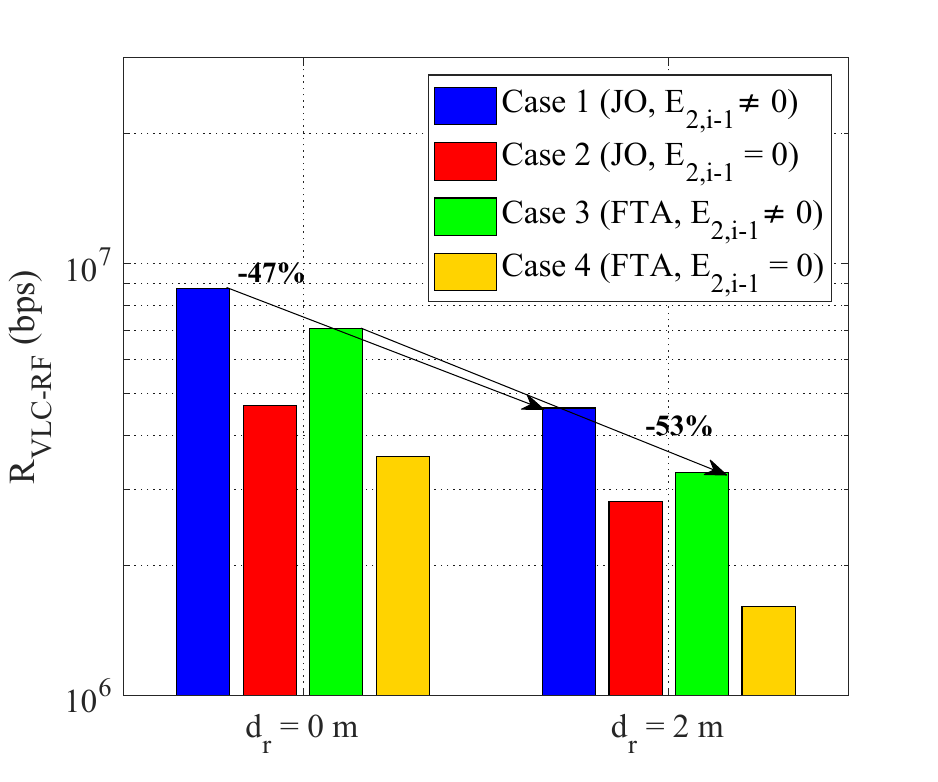} 
\caption{Optimal data rate}\label{fig:rate_random_5G} 
\end{subfigure}
\begin{subfigure}{0.32\linewidth} 
\centering
\includegraphics[trim=0.2cm 0.5cm 0.5cm 0.5cm, clip,width=\linewidth]{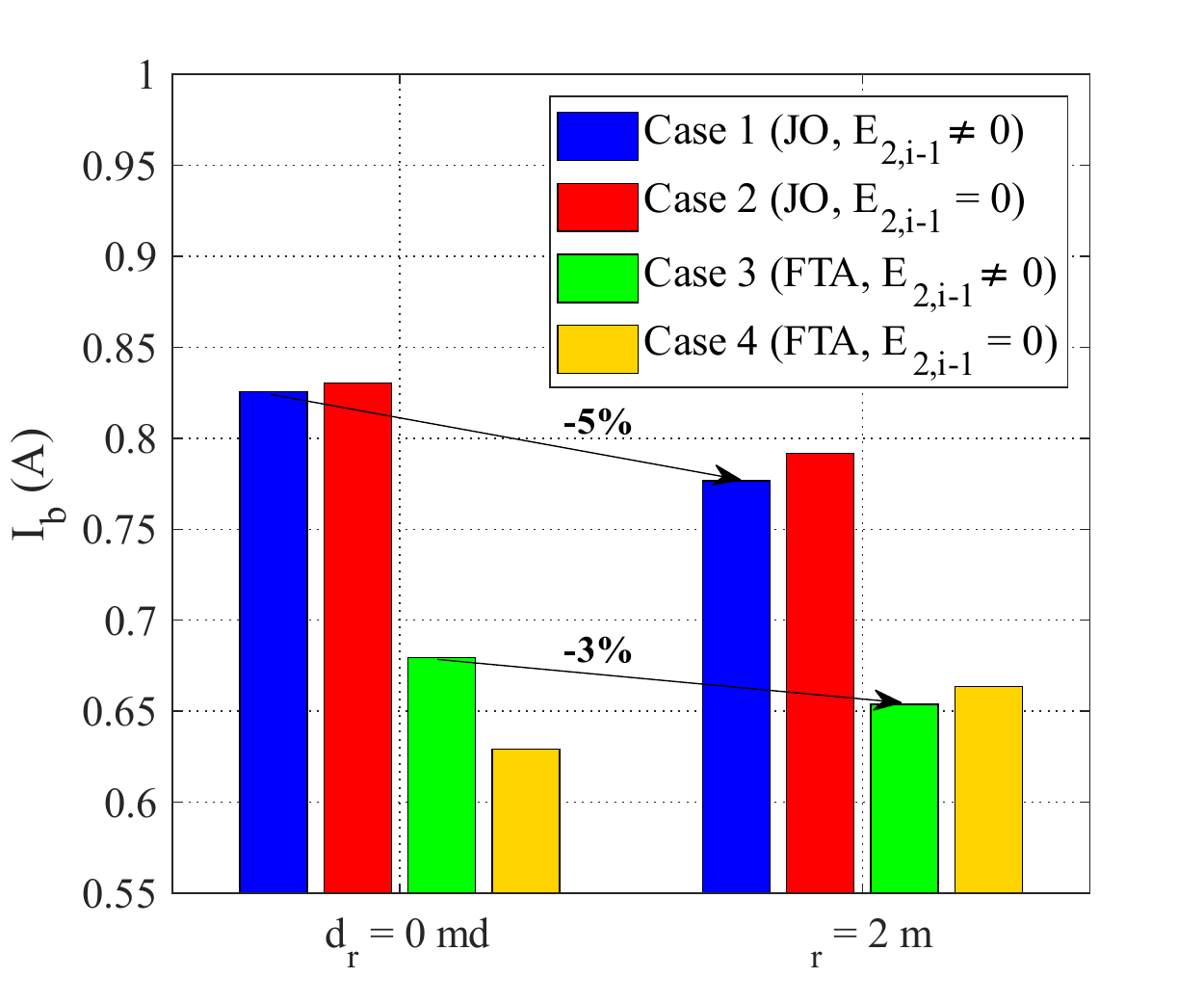} 
\caption{Optimal DC bias}
\label{fig:Ib_random_5G} 
\end{subfigure}
\begin{subfigure}{0.32\linewidth} 
\centering
\includegraphics[trim=0.2cm 0.5cm 0.5cm 0.5cm, clip,width=\linewidth]{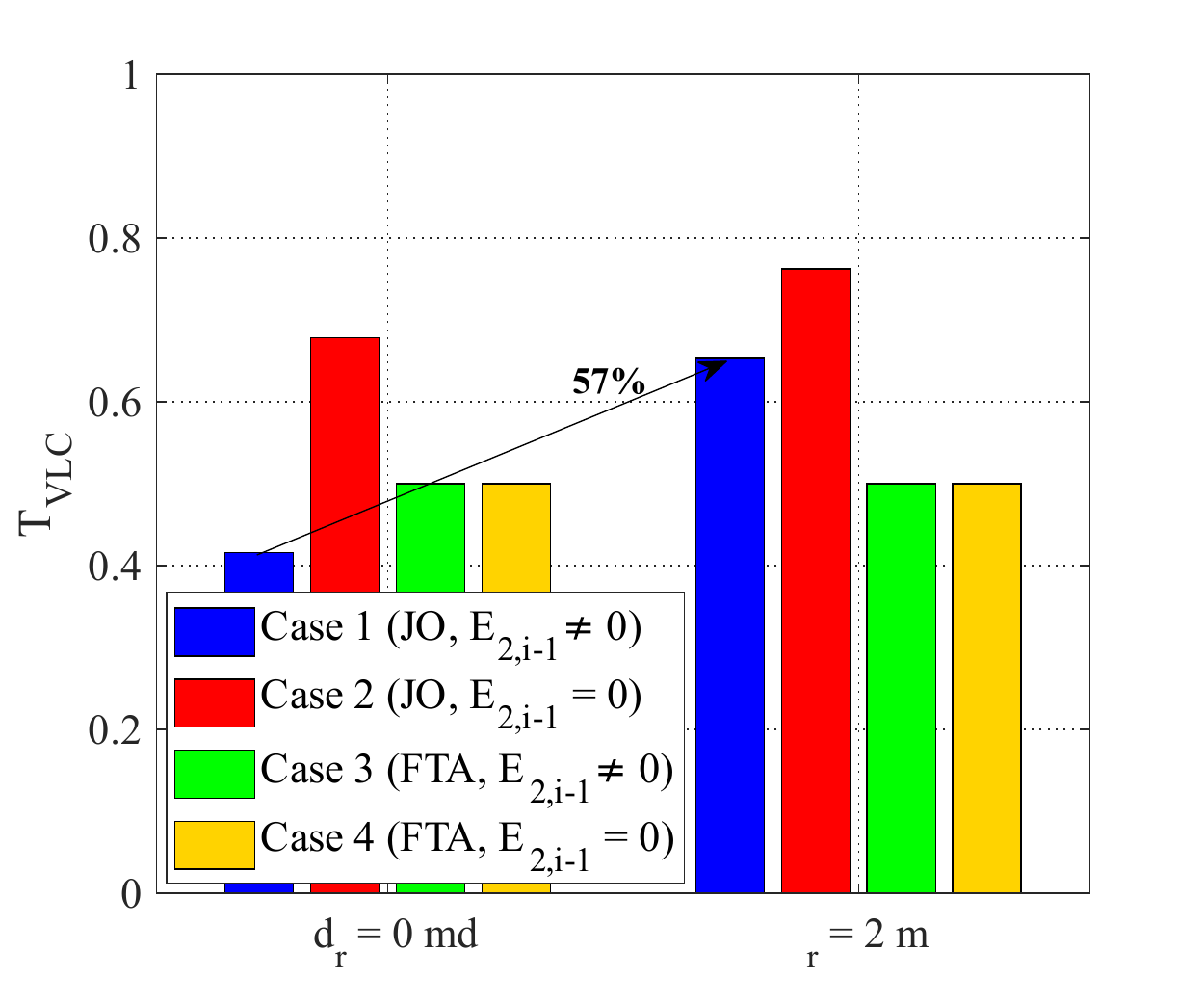} 
\caption{Optimal duration of VLC transmission}\label{fig:T_random_5G}
\end{subfigure} 
\caption{The performance of system under consideration with $f_{\textrm{c}} = 5$ GHz when the user node distance follows ${d_{\textrm{u}}} \sim \mathcal{U}\left[ {4, 8} \right]$.}
\label{fig:random_locations5G}
\end{figure*}

For both relay positions ($d_{\textrm{r}} = 0$~m and $d_{\textrm{r}} = 2$~m), the RF link serves as the bottleneck, as the achievable data rate is constrained by the lower data rate between the VLC and RF links (see~\eqref{eq:9}). Thus, the restriction in system performance is dominated by the RF link, especially as the user distance increases. Notably, this result demonstrates the critical advantage of leveraging harvested energy during RF transmissions, as proposed in our system design. It provides significant gains in the overall data rate by mitigating the limitations of the RF link.

Fig.~\ref{fig:random_locations2G} presents the performance of the system under consideration when the relay location is varied as $d_{\textrm{r}} \in \{0, 2\}$~m, and the user node distance follows a uniform distribution with ${{d_{\textrm{u}}} \sim \mathcal{U}[4, 8]}$~m. The RF frequency is assumed to be ${f_{\textrm{c}} = 2.4}$~GHz.
As observed in Fig.~\ref{fig:rate_random_2G},  harvesting energy during the RF transmission (Case 1 and Case 3) leads to a significant improvement in the optimal data rate. This performance boost is due to the ability of the relay to harvest energy during the RF phase, which subsequently powers the RF transmission.
To investigate the underlying factors contributing to this improvement, we further analyze the optimal DC bias and the time allocation for the VLC link in Fig.~\ref{fig:Ib_random_2G} and Fig.~\ref{fig:T_random_2G}, respectively.
Fig.~\ref{fig:Ib_random_2G} demonstrates that the optimal DC bias for the JO cases, regardless of energy harvesting during RF transmission, is higher than that for the FTA cases. This results in a lower peak amplitude of the input electrical signal, as highlighted in~\eqref{eq:13}, which ultimately constrains the VLC data rate. Conversely, increasing the DC bias results in more energy being harvested during VLC transmission (see~\eqref{eq:4}).
As shown in Fig.~\ref{fig:T_random_2G}, the optimal time allocated for VLC transmission, $T_{\textrm{VLC}}$, is less than 0.5 for Case 1 when $d_{\textrm{r}}$ = 0~m, which further restricts the VLC data rate (cf. \eqref{eq:2}). Although the difference in the optimal DC bias between Case 1 and Case 2 is practically negligible, the time allocated to VLC transmission in Case 2 is significantly larger than in Case 1. This is because in Case 2, the power for the relay depends entirely on the energy harvested during the VLC phase, as described in~\eqref{eq:4}, leading to more time being allocated to compensate for the lack of energy harvested during RF transmission.

\begin{figure}[t]
\centering
\begin{subfigure}{\linewidth} 
\centering
\includegraphics[width=\linewidth]{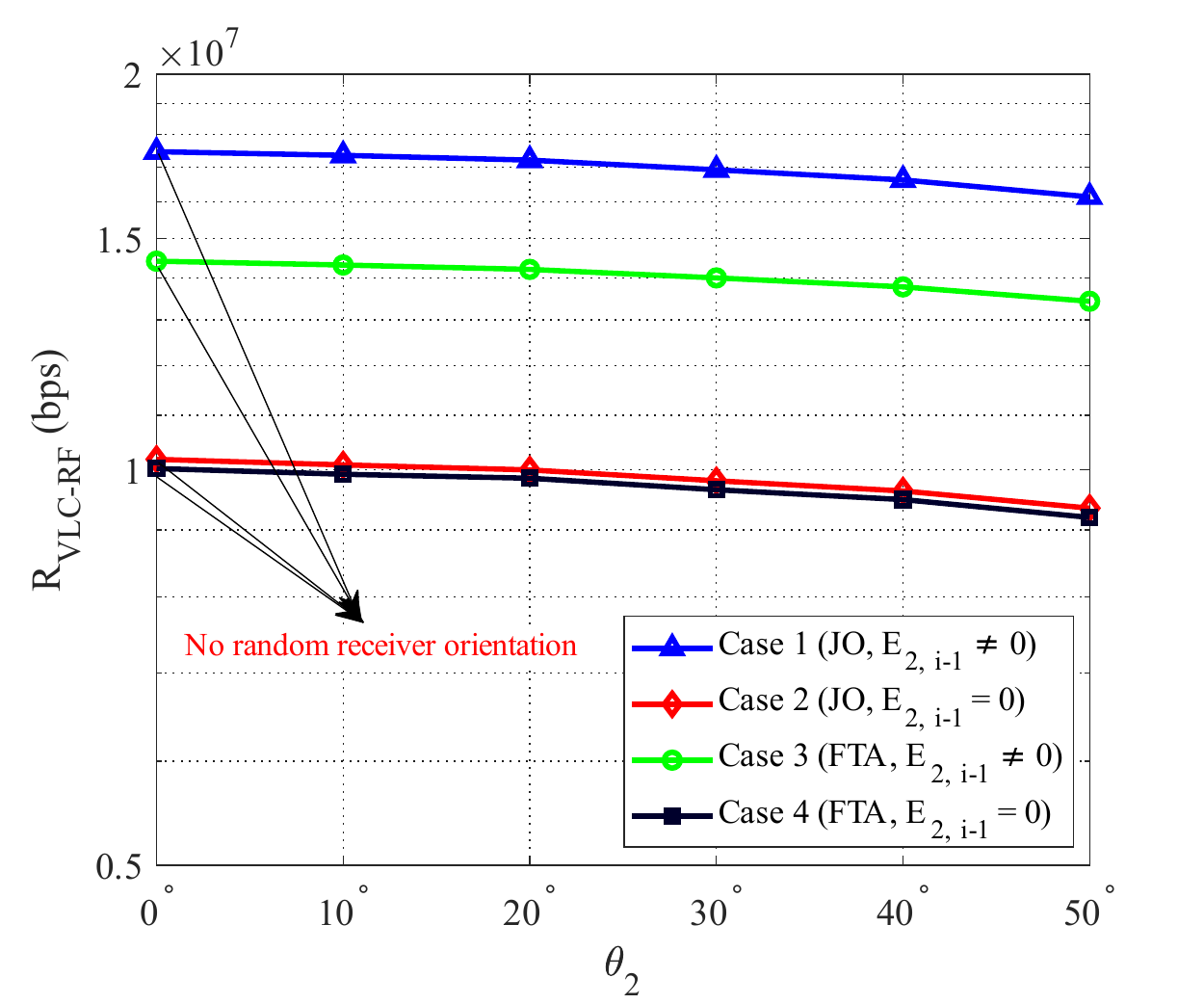} 
\caption{Relay location: $d_{\textrm{r}} = 0$~m.}\label{fig:dr_0} 
\end{subfigure}
\begin{subfigure}{\linewidth} 
\centering
\includegraphics[clip,width=\linewidth]{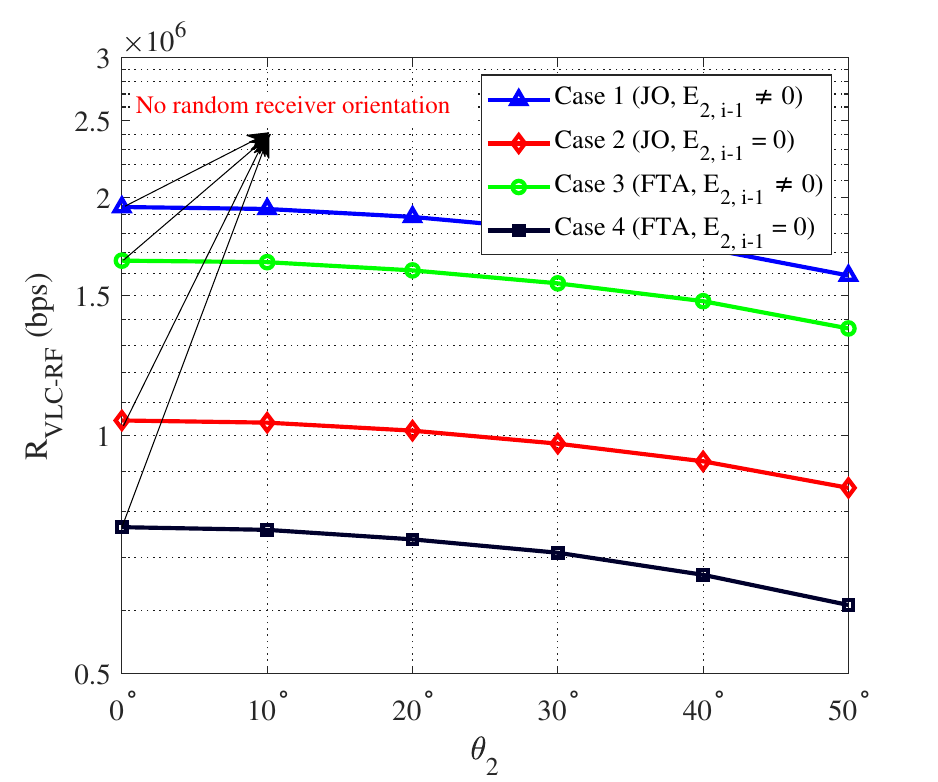} 
\caption{Relay location: $d_{\textrm{r}} = 4$~m.}
\label{fig:dr_4} 
\end{subfigure} 
\caption{The effect of random orientation on the optimal data rate of the RF user location follows a Uniform distribution with $\mathcal{U}\left[4, 8 \right]$ when the half-power beamwidth is $\Phi = 60^{\circ}$.  
}
\label{fig:random_theta}
\end{figure}

\begin{figure}[t]
\centering
\begin{subfigure}{\linewidth} 
\centering
\includegraphics[width=\linewidth]{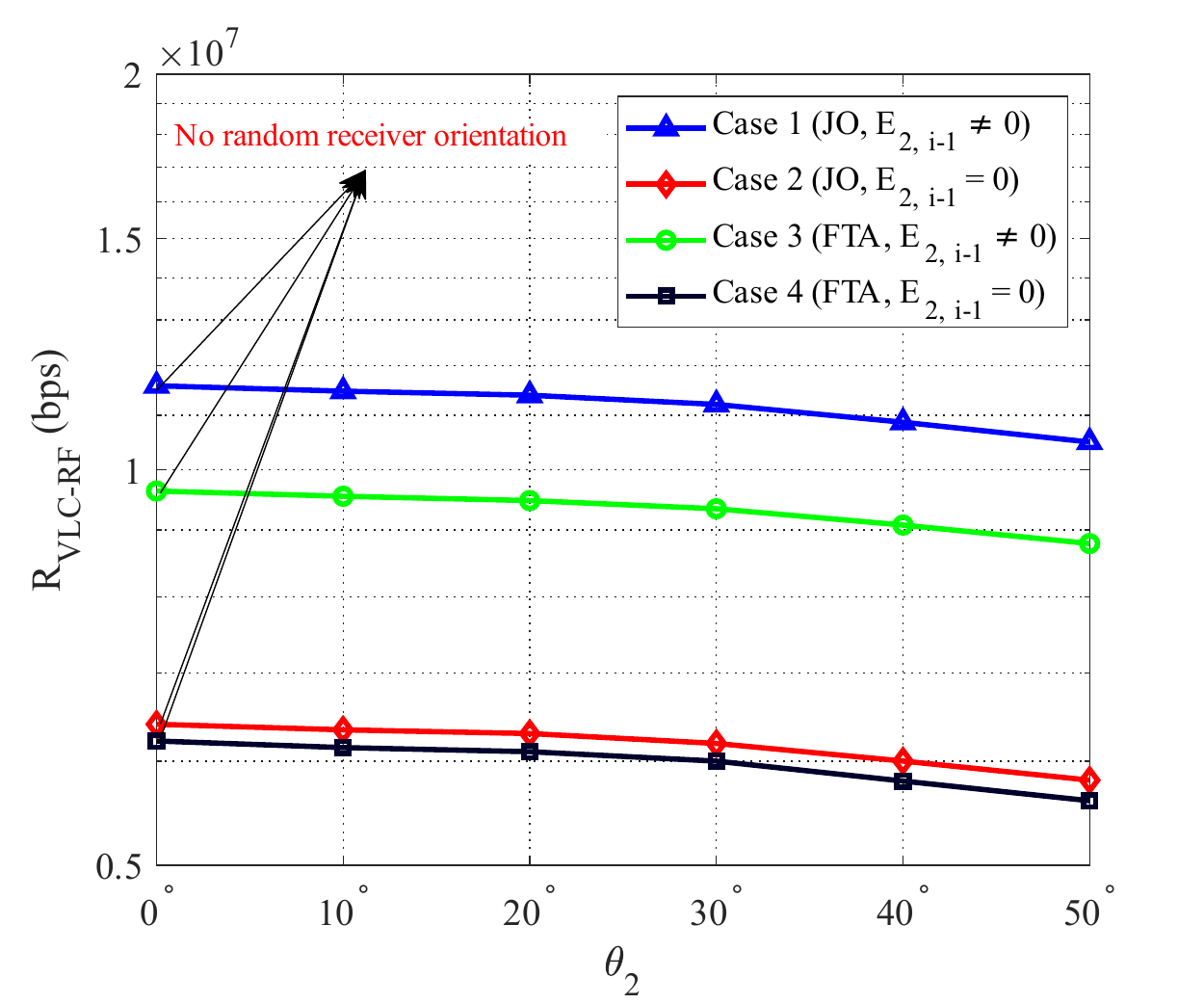} 
\caption{Relay location: $d_{\textrm{r}} = 0$~m.}\label{fig:dr_0_phi90} 
\end{subfigure}
\begin{subfigure}{\linewidth} 
\centering
\includegraphics[clip,width=\linewidth]{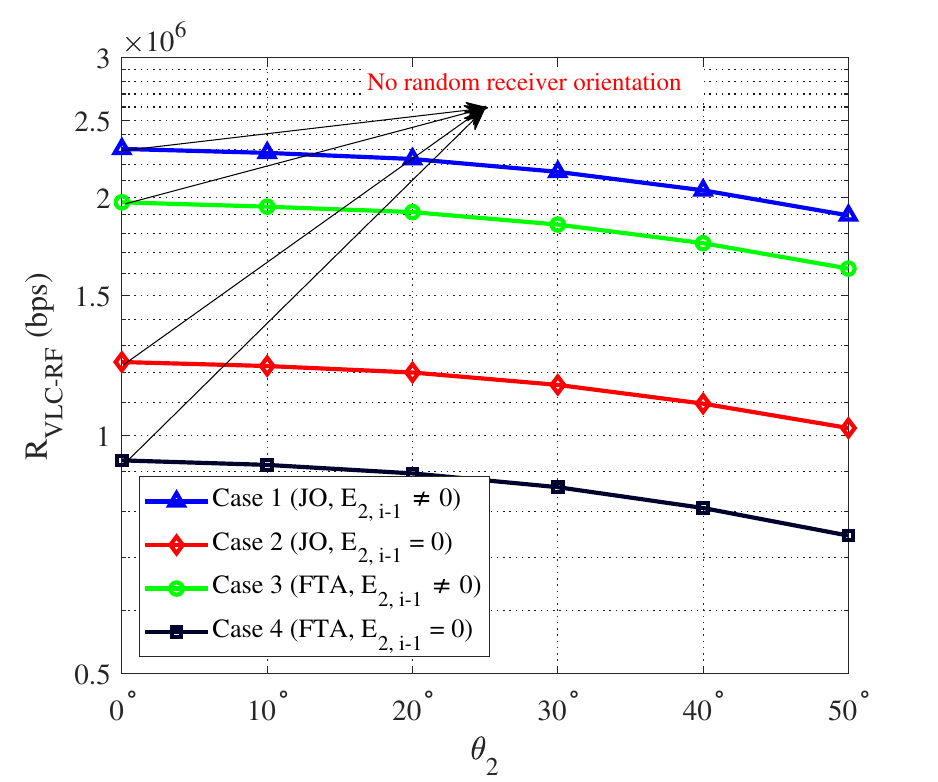} 
\caption{Relay location: $d_{\textrm{r}} = 4$~m.}
\label{fig:dr_4_phi90} 
\end{subfigure} 
\caption{The effect of random orientation on the optimal data rate of the RF user location follows a Uniform distribution with $\mathcal{U}\left[4, 8 \right]$ when the half-power beamwidth is $\Phi = 90^{\circ}$.  
}
\label{fig:random_theta_phi90}
\end{figure}

To study the effect of RF frequency, Fig.~\ref{fig:random_locations5G} presents the system performance for an RF carrier frequency of $f_{\textrm{c}} = 5$~GHz, assuming that the user node distance follows ${d_{\textrm{u}}} \sim \mathcal{U}[4, 8]$~m, and the relay location varies as $d_{\textrm{r}} \in \{0, 2\}$~m.
A comparison between Fig.~\ref{fig:rate_random_2G} and Fig.~\ref{fig:rate_random_5G} reveals that the optimal data rate decreases as the RF frequency increases. This behavior is expected, as the higher RF frequency leads to increased path loss, as described by~\eqref{eq:8}.
To further understand the behavior of the optimization, Figs.~\ref{fig:Ib_random_5G} and~\ref{fig:T_random_5G}  illustrate the optimal DC bias and the optimal time allocation for the VLC transmission, respectively. As shown in Fig.~\ref{fig:Ib_random_5G}, except for Case 4, the optimal DC bias decreases as the relay distance ($d_{\textrm{r}}$) increases, which results in an increase in the VLC link data rate. For Case 4, increasing the DC bias leads to greater energy harvesting during VLC transmission to compensate for the fact that the relay does not harvest energy during RF transmission. 
Comparing Figs.~\ref{fig:T_random_2G} and~\ref{fig:T_random_5G}, we observe that the time allocated to VLC transmission significantly increases as the carrier frequency increases to mitigate the higher path loss at 5~GHz to some extent. In particular, this has two key effects: 1) It increases the harvested energy during VLC transmission, and 2) It reduces the RF transmission time, thereby increasing the RF link power.

\subsection{Achievable Data Rate with Relay Random Orientation}\label{sec:num_results_ro}
In this subsection, we investigate the effect of relay random orientation on the maximum achievable data rate. Similar to the previous subsection, we consider the same transmission policies. Due to the complexity of the expressions for the average data rate of VLC and RF, the optimal values of $T_{\textrm{VLC}}$ and $I_{\textrm{b}}$ are obtained through exhaustive search. We assume the half-power beamwidth is $\Phi = 60^{\circ}$, the RF user location follows $d_{\textrm{u}} \sim U[4, 8]$~m, and the RF frequency is ${f_{\textrm{c}} = 2.4}$~GHz.
In Fig.~\ref{fig:dr_0}, the relay is located at $d_{\textrm{r}} = 0$~m, while in Fig.~\ref{fig:dr_4}, the relay distance is set to $d_{\textrm{r}} = 4$~m. The random orientation angle of the relay follows a uniform distribution $\theta_{\textrm{r}} \sim U[0, \theta_2]$, where $\theta_2$ varies within the range of $[0^{\circ}, 50^{\circ}]$.

As shown in Fig.~\ref{fig:random_theta}, the optimal data rate decreases as the range of random orientation increases. This reduction in performance can be attributed to the decrease in VLC channel gain. The decrease in channel gain directly impacts the amount of energy harvested by the relay, which in turn powers the RF link. The results also highlight the advantage of energy harvesting during RF transmission (Case 1 and Case 3), leading to improved data rates compared to other transmission policies.
Our proposed scheme outperforms the alternatives because it adjusts both the VLC link time duration ($T_{\textrm{VLC}}$) and the DC bias ($I_{\textrm{b}}$).

Next, in Fig.~\ref{fig:random_theta_phi90}, we investigate the effect of the Lambertian order by setting the half-power beamwidth to $\Phi = 90^{\circ}$. The directivity of a light source decreases as the half-power beamwidth increases. As shown in Fig.~\ref{fig:dr_0_phi90}, when the relay is at $d_{\textrm{r}} = 0$~m, the maximum achievable data rate decreases for the larger half-power beamwidth case. As the distance between the LED and relay increases, the received power decreases due to the spreading of the emitted light. However, this reduction can be mitigated by adjusting the Lambertian order, which helps distribute power more evenly over a broader range of angles. The advantage of selecting an appropriate half-power beamwidth becomes more apparent as the relay distance increases. For example, the comparison between Fig.~\ref{fig:dr_4} and Fig.~\ref{fig:dr_4_phi90} demonstrates that the achievable data rate increases with a larger half-power beamwidth.

\section{Conclusion}\label{sec:conclusion}
In this paper, we proposed a joint optimization framework for energy-harvesting hybrid VLC-RF networks, designed to maximize data rate performance by optimizing both the DC bias and the time allocated for VLC transmission. Our approach allows the relay to harvest energy during both VLC transmission and RF communication, enabling more efficient energy utilization. By dividing the optimization problem into two subproblems, we addressed the non-convex DC bias issue through the MM approach, while optimizing the VLC transmission time in the second step. The results demonstrated that this joint optimization approach significantly outperforms methods that optimize only one parameter, such as the DC bias, providing superior data rates across a variety of operating conditions. These improvements were especially noticeable in scenarios involving greater relay distances and higher RF frequencies, where the system exhibited robust performance despite the additional challenges. We also examined the effects of random receiver orientation, noting that its impact grows as the relay distance increases. Additionally, we showed that adjusting the half-power beamwidth plays a crucial role in maintaining data rates, with larger beamwidths mitigating performance losses at greater distances, even if they initially reduce the data rate at shorter distances.
Our findings have significant implications for the design of energy-efficient hybrid communication systems, particularly in dense indoor environments where spectrum resources are limited. By effectively managing energy harvesting and transmission parameters, our framework enhances system performance and reliability.

Future work could extend this framework to handle multi-antenna and multi-user networks, enabling more efficient energy harvesting and directional transmission. Furthermore, integrating machine learning techniques for real-time optimization could allow the system to dynamically adapt to changing environments and user demands. Addressing practical issues such as channel variability, blockages, and cooperative communication strategies could further enhance the system's performance in real-world applications, such as smart homes and industrial IoT networks.

\bibliographystyle{unsrt}
\bibliography{references}

\begin{IEEEbiography}[{\includegraphics[width=1in,height=1.25in,clip,keepaspectratio]{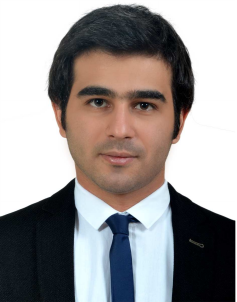}}]{Amir Hossein Fahim Raouf}
received his B.Sc. degree in Electrical and Computer Engineering from Isfahan University of Technology, Isfahan, Iran, and his M.Sc. degree in Electrical and Electronics Engineering from Ozyegin University, Istanbul, Turkey. He is currently pursuing a Ph.D. degree in Electrical Engineering at North Carolina State University. His main research areas are wireless communications, visible light communications, and quantum key distribution.
\end{IEEEbiography}
\begin{IEEEbiography}[{\includegraphics[width=1in,height=1.25in,clip,keepaspectratio]{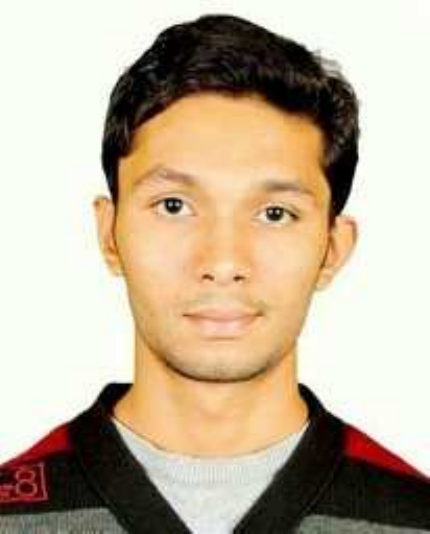}}]{Chethan Kumar Anjinappa}
received a B.E. degree in electronics and communication engineering from the Sri Jayachamarajendra College of Engineering, Mysuru, India, in 2012 and an M.E. degree in signal processing from the Indian Institute of Science (IISc), Bengaluru, India, in 2016. He received his Ph.D. degree in electrical engineering from North Carolina State University, Raleigh, NC, USA. He currently works at Ericsson Research, CA, USA. His research interests include 5G and mmWave communication, physical layer security, sparse signal processing, and machine learning.
\end{IEEEbiography}
\begin{IEEEbiography}[{\includegraphics[width=1in,height=1.25in,clip,keepaspectratio]{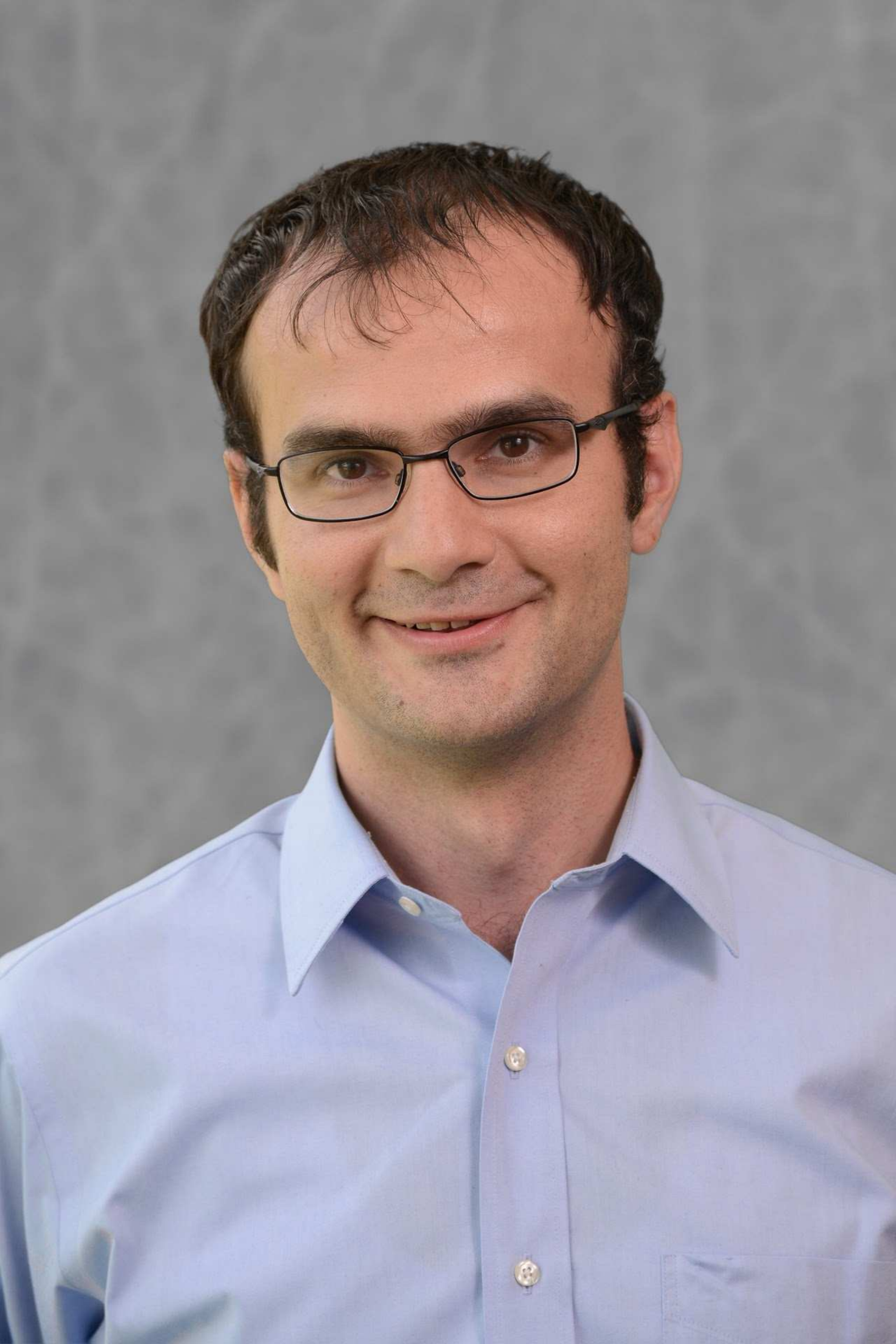}}]{Ismail Guvenc}
(Fellow, IEEE) received his Ph.D. degree in electrical engineering from the University of South Florida in 2006. He was with Mitsubishi Electric Research Labs during 2005, with DOCOMO Innovations between 2006-2012, and with Florida International University between 2012-2016. He is a Professor at the Department of Electrical and Computer Engineering at North Carolina State University. His recent research interests include 5G wireless systems, communications and networking with drones, and heterogeneous wireless networks. He has published more than 200 conference/journal papers and book chapters, and several standardization contributions. He co-authored/co-edited three books for Cambridge University Press, served as an editor for IEEE Communications Letters (2010-2015), IEEE Wireless Communications Letters (2011-2016), IEEE Transactions on Wireless Communications (2016-present), and IEEE Transactions on Communications (2020-present), and as a guest editor for several other journals. Dr. Guvenc is an inventor/coinventor in some 30 U.S. patents and he is a senior member of the National Academy of Inventors. He is a recipient of the University Faculty Scholar Award (2021), NCSU ECE R. Ray Bennett Faculty Fellow Award (2019), FIU College of Engineering Faculty Research Award (2016), NSF CAREER Award (2015), Ralph E. Powe Junior Faculty Enhancement Award (2014), and USF Outstanding Dissertation Award (2006). 
\end{IEEEbiography}

\EOD
\end{document}